\documentclass[%
reprint,
 amsmath,amssymb,
 aps,
floatfix,
]{revtex4-2}

\usepackage{xcolor}
\usepackage{graphicx}
\usepackage{dcolumn}
\usepackage{bm}
\usepackage{hyperref}
\usepackage{physics}
\usepackage{mathtools}
\usepackage{enumitem}
\usepackage{subeqnarray}

\newcommand{\conj}[1]{#1^*}
\newcommand{\imag}[1]{\text{Im}\left[#1\right]}

\newcommand{\opf}{\,\hat{f}}
\newcommand{\opfd}{\,\hat{f}\,^\dagger}

\newcommand{\opg}{\,\bar{E}}

\newcommand{\epspp}{\varepsilon''}
\newcommand{\pdt}{\,\partial_t}
\newcommand{\chit}{\chi^{(2)}}

\renewcommand{\exp}[1]{\,e^{#1}}

\newcommand{\pe}{\hat{E}^{(+)}}
\renewcommand{\ne}{\hat{E}^{(-)}}
\newcommand{\pme}{\hat{E}^{(\pm)}}
\newcommand{\pump}[1]{E^{(+)}_{P,#1}}
\newcommand{\pumpn}[1]{E^{(-)}_{P,#1}}
\newcommand{\fpump}[1]{\mathcal{E}^{(+)}_{P,#1}}
\newcommand{\fpumpn}[1]{\mathcal{E}^{(-)}_{P,#1}}

\newcommand{\epss}{\varepsilon_0}

\let\vec\mathbf

\newcommand{\rr}{\mathbf{r}}
\newcommand{\xx}{\bm{\xi}}
\newcommand{\XX}{\bm{\Xi}}

\begin{document}

\preprint{APS/123-QED}

\title{
Non-perturbative theory of spontaneous parametric down-conversion in open and dispersive optical systems
}

\author{Aleksa Krsti\' c}
\email{aleksa.krstic@uni-jena.de}
\affiliation{
	Institute of Applied Physics, Abbe Center of Photonics, Friedrich-Schiller University Jena, Albert-Einstein-Straße 15, 07745 Jena, Germany}%
 
\author{Frank Setzpfandt}\altaffiliation[Also at ]{Fraunhofer Institute for Applied Optics and Precision Engineering, Albert-Einstein-Straße 7, 07745 Jena, Germany} 
\affiliation{
	Institute of Applied Physics, Abbe Center of Photonics, Friedrich-Schiller University Jena, Albert-Einstein-Straße 15, 07745 Jena, Germany}%

\author{Sina Saravi}
\affiliation{
 Institute of Applied Physics, Abbe Center of Photonics, Friedrich-Schiller University Jena, Albert-Einstein-Straße 15, 07745 Jena, Germany}%

\date{\today}

\begin{abstract}
We develop a non-perturbative formulation based on the Green-function quantisation method, that can describe spontaneous parametric down-conversion in the high-gain regime in nonlinear optical structures with arbitrary amount of loss and dispersion. This formalism opens the way for description and design of arbitrary complex and/or open nanostructured nonlinear optical systems in quantum technology applications, such as squeezed-light generation, nonlinearity-based quantum sensing, and hybrid quantum systems mediated by nonlinear interactions. As an example case, we numerically investigate the scenario of integrated quantum spectroscopy with undetected photons, in the high-gain regime, and uncover novel gain-dependent effects in the performance of the system.
\end{abstract}

\maketitle

\arraycolsep 0.05em
\section{\label{sec:intro}Introduction}
With the ever growing importance of optical quantum technologies, new ways to leverage nonclassical properties of light are in high demand \cite{obrien2009photonic,wang2020integrated-quantum}. Some of the most prevalent methods for generating non-classical light are based on nonlinear sources of photon pairs in which nonlinear optical processes, such as spontaneous parametric down-conversion (SPDC) or spontaneous four-wave mixing (SFWM), are used to convert input light into pairs of photons which exhibit non-classical correlations \cite{boyd2020nonlinear,caspani2017integrated-nonlinear}, highly relevant for many applications in quantum technologies. Such sources also offer the advantage that they can be implemented in integrated platforms \cite{wang2020integrated-quantum,caspani2017integrated-nonlinear}, as well as utilised in hybrid quantum photonic systems, to overcome the limitations of monolithic integrated systems \cite{elshaari2020hybrid}.

Apart from generating entangled photon pairs, nonlinear photon-pair sources are also commonly used as a means to obtain heralded single photons \cite{anwar2021entangled,eisaman2011nonl-single-ph,meyer2020single}. In both cases the sources are operated in the so-called low-gain regime, where the dominant contribution to the output state (apart from the vacuum) is a single photon-pair state \cite{eisaman2011nonl-single-ph}. This is achieved by keeping the input pump beam of the nonlinear source at a power that is low enough to keep the multiple photon-pair contributions in the output state negligible. Although such low-gain probabilistic sources of photon pairs are of importance in many areas of quantum technologies, their development in the high-gain regime is also of interest \cite{quesada2022beyond}. In the high-gain regime, the SPDC and SFWM sources can be used as sources of squeezed light, where squeezed light has wide applications in continuous variable (CV) quantum computation \cite{menicucci2006cv-computing,larsen2021fault-cv-computing,zhong2020quantum-advantage,arrazola2021quantum}, quantum communication \cite{vanLook2002squeezedCommunication}, as well as quantum sensing \cite{cutipa2022bsv-sensing,panahiyan2022tpa-squeezed,lawrie2019squeezedSensing}. Another application of high-gain operation of such sources can be for generating multi-photon Fock states \cite{engelkemeier2020multiFock,engelkemeier2021climbing}, which have applications in many branches of quantum technology as well as metrology and fundamental tests of quantum entanglement \cite{dell2006multiphoton,holland1993interferometric,sperling2019mode-entanglement-metrology}.

The rapidly growing interest in the utilisation of high-gain nonlinear sources of quantum light in nanostructured and hybrid platforms \cite{arrazola2021quantum, madsen2022quantum, vaidya2020broadband, nehra2022few, chen2022ultra, peace2022picosecond,stokowski2022integrated} necessitates the development of new theoretical formalisms to describe high-gain SPDC and SFWM, as such systems can be generally open/lossy with highly complex spatial and spectral properties \cite{busch2007periodic-nano,lalanne2018resonant}.
Formalisms for describing SPDC and SFWM in the high-gain regime in closed systems are well-established, but either neglect loss altogether \cite{wasilewski2006pulsed,christ2013theory,chekhova2020BSV-increasingBrightness} or are only able to account for weak losses that have negligible effect on the modal structure of the system \cite{helt2020degenerate,quesada2022beyond}, but are not capable of treating systems where loss is an inherent part of the system's optical properties, such as nanoresonators \cite{gigli2020quasinormal}, plasmonic systems \cite{lalanne2018resonant}, or structures with inherently leaky guided modes \cite{lecamp2007theoretical,saravi2015phase}, or strongly lossy systems that can be encountered in quantum sensing applications \cite{kumar2020integrated}. Finally, many established high-gain formalisms use a weak-dispersion approximation for the involved modes or rely on approximate expansions to accommodate higher-order dispersions \cite{christ2013theory,quesada2022beyond,triginer2020twin-beam}, and do not offer a straightforward path to inclusion of complex dispersion relations, such as those that can appear in photonic crystals with internal loss or gain, where the combination of loss/gain and evanescent modes create complex dispersion diagrams \cite{grgic2012fundamental,saravi2016effect}.

The Green's function (GF) quantisation method \cite{vogel2006quantum,gruner1996gf-quant,dung1998gf-quant} offers a way to describe light quantisation in optical systems with arbitrary dispersion, loss, and nanostructuring, as the classical GF of the system incorporated into this quantisation approach naturally takes all these effects into account. In contrast to closed systems, where the normal modes of the field are quantised to obtain photon creation/annihilation operators, the GF method quantises the local bosonic excitations of the combined field and medium system, which allows it to naturally describe any form of loss as a coupling between different field-matter modes. A detailed overview of the GF quantisation procedure can be found in Ref.~\cite{knoll2003qed}. The GF formalism has already been used to describe photon-pair generation in plasmonic structures \cite{andrey2016prl} and dielectric nanoresonators and nanoparticles \cite{olekhno2018spontaneous,nikolaeva2021directionalGF,poddubny2018nonlinear}, as well as in the presence of quantum emitters and significant loss \cite{saravi2017amspdc,kumar2020integrated, santos2022subdiffraction}, but in all these cases, it has only been used do describe pair generation in the low-gain regime. 

In this work, we present a formalism, based on the GF quantisation method, to describe high-gain SPDC in arbitrary open and dispersive optical systems, within the undepleted pump approximation.
Our developed non-perturbative formalism allows the exact calculation of the field operators, while intrinsically taking into account arbitrary dispersion and losses of nanostructured systems through the classical electromagnetic GF of the optical system.
This method can be of special interest for the description of engineered squeezed-light generation in nanostructured systems in CV quantum computing applications \cite{arrazola2021quantum} or systems where losses cannot be treated perturbatively (e.g. nonlinear metasurfaces for quantum light generation \cite{chekhova2020quantumMetasurfaces,santiago2021photon,zhang2022spatially}), as well as high-gain quantum sensing and imaging applications, where loss is not necessarily a weak effect \cite{kumar2020integrated,santos2022subdiffraction,cutipa2022bsv-sensing,panahiyan2022tpa-squeezed}. Additionally, our formalism may open the path for the high-gain description of hybrid nonlinear systems, which involve direct interfacing of quantum emitters with nonlinear systems \cite{saravi2017amspdc}.

The paper is organised as follows: In Sec.~\ref{sec:theory1} we introduce the theoretical framework for describing SPDC in arbitrarily dispersive and lossy systems using fields quantised using the GF quantisation method. We derive coupled-mode equations that allow the calculation of the full electric field operator at arbitrary times and amounts of gain (within the undepleted pump approximation). Then, we introduce a modified formalism for the calculation of frequency-domain field components, which is extremely useful in evaluating the spectral properties of the output quantum state. In Sec.~\ref{sec:qsup}, as an example application scenario, we apply the formalism to the problem of integrated quantum spectroscopy with undetected photons (QSUP) and investigate the effects of a spectrally localised loss on the spectrum of the output photons of a waveguide SPDC source, operating in the high-gain regime. We show that the performance of the sensing system is gain-dependent and its signatures become more prominent as gain is increased, but also that they saturate at even higher gain values. Finally, in Sec.~\ref{sec:conclusion} we discuss our results and review the advantages and use cases of our non-perturbative formalism.

\section{\label{sec:theory1}Non-perturbative description of SPDC in open and dispersive systems}
We are considering a generally inhomogeneous, dispersive and lossy dielectric system, characterised by the linear permittivity $\varepsilon(\rr,\omega)=\varepsilon'(\rr,\omega)+i\epspp(\rr,\omega)$, where $\varepsilon'$ and $\epspp$ are its real and imaginary parts, respectively. To make our expressions less cumbersome, we restrict our present analysis to an isotropic system and note that a generalisation to anisotropic systems is possible by using an appropriately generalised GF quantisation procedure, such as the one derived in Ref.~\cite{suttorp2007anisotropGF}. 

We consider the system to have a second-order nonlinearity, characterised by the second-order nonlinear susceptibility tensor $\chit_{ijk}(\rr)$. In the Heisenberg picture, the SPDC interaction Hamiltonian has the following form \cite{sharapova2015schmidtModes}:
\begin{eqnarray}\label{eq:hamiltonian}
\hat{H}_\mathrm{SPDC}(t)&=&-\epss\int \dd{\rr}\chi^{(2)}_{ijk}(\rr)\pumpn{i}(\rr,t) \nonumber\\
 &&\times \pe_j (\rr,t)\pe_k (\rr,t) \nonumber\\ 
 &&+H.c.,
\end{eqnarray}
where $\pe_{j,k} (\rr,t)$ are Cartesian components of the positive-frequency part of the generated quantum fields, $\pumpn{i}(\rr,t)$ are the Cartesian components of the negative-frequency part of the pump field and $\epss$ is the dielectric permittivity of vacuum. We emphasise that we are working in the undepleted pump approximation and thus assume the pump field to be an undepleted classical pulse, which results in the Hamiltonian being quadratic in the field operators and is a crucial assumption for finding the input-output operator relations later in this work. Treating the extremely high-gain scenarios, in which the pump can also deplete substantially or ones where the pump is in a few-photon state at the input, results in more complex dynamics which cannot be described by the Hamiltonian of the quadratic form given in Eq.~\eqref{eq:hamiltonian}.
However, such scenarios can be used for generation of non-Gaussian quantum states \cite{yanagimoto2022onset,xing2023pump} and advancing the GF formalism to go beyond the undepleted pump approximation could be an interesting direction for future works. 

In addition, we note that for our calculations, we assume a non-dispersive and real-valued $\chit$ nonlinearity, which corresponds to operating the nonlinear material in spectral regions that are far from its material resonances. Such an approximation is widely used in treatment of high-gain squeezed-light generation in most scenarios of practical interest \cite{quesada2022beyond,wasilewski2006pulsed,christ2013theory,chekhova2020BSV-increasingBrightness,helt2020degenerate,sharapova2015schmidtModes,quesada-sipe2020twinBeam-waveguides}. In principle, our approach could be generalized to treat dispersive nonlinearities, by starting the calculations from a more complex nonlinear Hamiltonian that includes a dispersive $\chit$ coefficient \cite{scheel2006nonlNoisePRL,scheel2006causalNonlinearQuantum,lindel2020polaritonicVacuum,lindel2021theoryPolaritonicVacuum}, provided that the Hamiltonian remains quadratic in the electric field operators. We also emphasize, that in our model, we assume an undepleted pump approximation, which is a well-justified approximation to almost all cases with a practically reachable value for the nonlinear gain. In this approximation, the nonlinear Hamiltonian keeps its quadratic form, allowing us to formulate an input-output relation for the electric field operator. Finally, it should be noted, that we do not consider the effect of nonlinear noise, predicted to have an effect for very strong pump intensities \cite{scheel2006nonlNoisePRL,scheel2006causalNonlinearQuantum}.

As will be the convention in the remainder of the paper, repeated Latin indices are implicitly summed over and the domains of integration for spatial and frequency integrals are $(-\infty,\infty)$ and $[0,\infty)$, respectively, unless otherwise noted in the text. We carry out all of our calculations in the Heisenberg picture and assume that the nonlinear interaction is adiabatically "turned on" at $t\rightarrow-\infty$ and adiabatically "turned off" as $t\rightarrow\infty$.

To allow for the treatment of open and dispersive systems, the electric field operator is quantised using the GF quantisation method \cite{vogel2006quantum,gruner1996gf-quant,dung1998gf-quant}:
\begin{eqnarray}
    \vec{\hat{E}}^{(\pm)}(\rr,t)&=&\int\dd{\omega}\vec{\underline{\hat{E}}}^{(\pm)}(\rr,\omega,t),\nonumber\\
    \underline{\hat{E}}^{(+)}_i(\rr,\omega,t)&=& i\mathcal{K}\,\frac{\omega^2}{c^2}\int\dd{\rr'}\sqrt{\epspp(\rr',\omega)}\nonumber\\
    && \times \,G_{ij}(\rr,\rr',\omega)\opf_j(\rr',\omega,t), \label{eq:gDefNormal}\\
    \underline{\hat{E}}^{(-)}_i(\rr, \omega,t)&=&\left(\underline{\hat{E}}^{(+)}_i(\rr,\omega,t)\right)^\dagger,\nonumber
\end{eqnarray}
where $\mathcal{K}=\sqrt{\frac{\hbar}{\pi \epss}}$, $\vec{\underline{\hat{E}}}^{(\pm)}(\rr,\omega,t)$ is the amplitude operator of frequency $\omega$, $G_{ij}(\rr,\rr',\omega)$ are the matrix elements of the dyadic GF of the dielectric system $\overleftrightarrow{\vec{G}}(\rr,\rr',\omega)$ and $\opf_j(\rr',\omega,t)$ are the Cartesian components of the bosonic annihilation operator $\vec{\opf}(\rr',\omega,t)$. 

The GF in \eqref{eq:gDefNormal} is the solution to the classical Helmholtz equation:
\begin{equation}\label{eq:gfHelmholtz}
\left(\vec{\nabla}\times\vec{\nabla}\times-\frac{\omega^2}{c^2}\varepsilon(\rr,\omega)\right)\overleftrightarrow{\vec{G}}(\rr,\rr',\omega)=\overleftrightarrow{I}\delta(\rr-\rr'),
\end{equation}
together with the condition that it vanishes at infinity. In the above equation $\overleftrightarrow{I}$ is the identity matrix.

The operator $\vec{\opf}(\rr',\omega,t)$ annihilates a local field-matter excitation of frequency $\omega$, located at $\rr'$. Along with its adjoint $\vec{\opf}^\dagger(\rr',\omega,t)$, it makes up the fundamental operator algebra of the GF quantisation method and obeys the canonical commutation relation:
\begin{equation}\label{eq:commf}
\comm{\opf_i(\rr,\omega,t)}{\opfd_j(\rr',\omega',t)}=\delta_{ij}\delta(\rr-\rr')\delta(\omega-\omega').
\end{equation}
Using Eq.~\eqref{eq:commf}, it can be shown that the amplitude operators $\vec{\underline{\hat{E}}}^{(\pm)}(\rr,\omega,t)$ obey the following commutation relation:
\begin{eqnarray}\label{eq:commgNormal}
    \comm{\underline{\hat{E}}_i^{(+)}(\rr,\omega,t)}{\underline{\hat{E}}_j^{(-)}(\rr',\omega',t)}&=&\mathcal{K}^2\frac{\omega^2}{c^2}\delta(\omega-\omega')\nonumber\\
    &&\times\imag{G_{ij}(\rr,\rr',\omega)},\quad
\end{eqnarray}
where $\imag{G_{ij}(\rr,\rr',\omega)}$ is the imaginary part of the Green's function. The full derivation of the above relation can be found in Appendix~\ref{app:commRelation}.

Before proceeding, we note that the explicit results of our non-perturbative approach, presented in this paper, are strictly valid for fields in \textit{non-magnetic systems}. However, the main ideas and steps of our formalism could potentially be adapted for magnetic systems by using an appropriate field quantisation, such as one used in Ref.~\cite{knoell2003gfMagnetic}.

\subsection{\label{sec:IOrelations}Equations of motion}
The Cartesian components of $\vec{\underline{\hat{E}}}^{(\pm)}(\rr,\omega,t)$ evolve in time according to the Heisenberg equation of motion:
\begin{equation}\label{eq:heisenbergEquation}
\pdt \underline{\hat{E}}_i^{(\pm)}(\rr,\omega,t)=\frac{1}{i \hbar}
\comm{\underline{\hat{E}}_i^{(\pm)}(\rr,\omega,t)}{\hat{H}(t)},
\end{equation}
where $\hat{H}(t)=\hat{H}_0(t)+\hat{H}_\mathrm{SPDC}(t)$ and $\hat{H}_0(t)=\hbar \int\dd{\rr}\int\dd{\omega}\omega \opfd_i(\rr,\omega,t)\opf_i(\rr,\omega,t)$ is the free-field Hamiltonian in the GF quantisation scheme \cite{gruner1996gf-quant}.
The resulting differential equations thus contain terms describing both the nonlinear and free-field evolution. The latter can be eliminated and the equations of motion made simpler by introducing the slowly-varying, \textit{rotating-frame} operators, $\vec{\opg}^{(\pm)}(\rr,\omega,t)$, where:
\begin{subeqnarray}
    \vec{\underline{\hat{E}}}^{(\pm)}(\rr,\omega,t)&=&\vec{\opg}^{(\pm)}(\rr,\omega,t)\exp{\mp i \omega t},\slabel{eq:slowlyVaryingSub}\\
    \vec{\hat{E}}^{(\pm)}(\rr,t)&=&\int\dd{\omega}\vec{\opg}^{(\pm)}(\rr,\omega,t)\exp{\mp i \omega t}.\slabel{eq:EwithSlowOps}
\end{subeqnarray}
The exponential factors account for the free-field component of the evolution and the rotating-frame operators $\vec{\opg}^{(\pm)}(\rr,\omega,t)$ evolve according to:
\begin{equation}\label{eq:rotFrameEquation}
\pdt \vec{\opg}^{(\pm)}(\rr,\omega,t)=
\frac{1}{i \hbar}\comm{\vec{\opg}^{(\pm)}(\rr,\omega,t)}{\hat{H}_\mathrm{SPDC}(t)}.
\end{equation}
A proof of the above equation, as well as a more formal definition of the rotating-frame operators is given in Appendix~\ref{app:rotatingFrame}.

Rotating-frame creation and annihilation operators $\bar{\vec{f}}^{(\dagger)}(\rr,\omega,t)$ are defined equivalently to Eq.~\eqref{eq:slowlyVaryingSub} and are related to the rotating frame amplitude operators $\vec{\opg}^{(\pm)}(\rr,\omega,t)$ by the relation:
\begin{eqnarray}
    \opg_i^{(+)}(\rr, \omega,t)&=&i\mathcal{K}\,\frac{\omega^2}{c^2}\int \dd{\rr'}\sqrt{\epspp(\rr',\omega)} \nonumber\\
    &&\times G_{ij}(\rr,\rr',\omega)\bar{f}_j(\rr',\omega,t),\quad\label{eq:gDef}
\end{eqnarray}
and its adjoint for $\opg_i^{(-)}(\rr, \omega,t)$ and $\bar{f}^\dagger_j(\rr',\omega,t)$.
The rotating-frame amplitude operators $\vec{\opg}^{(\pm)}(\rr,\omega,t)$, as well as the rotating-frame creation/annihilation operators $\vec{\bar{f}}^{(\dagger)}(\rr',\omega,t)$, can be shown to obey equivalent commutation relations as their Heisenberg picture counterparts:
\begin{subeqnarray}
\left[\bar{f}_i(\rr,\omega,t),\bar{f}^{\dagger}_j(\rr',\omega',t)\right]=
\delta_{ij}&\delta&(\rr-\rr')\delta(\omega-\omega'),\qquad\slabel{eq:commFbar}\\
    \comm{\opg_i^{(+)}(\rr,\omega,t)}{\opg_j^{(-)}(\rr',\omega',t)}&=& \mathcal{K}^2\frac{\omega^2}{c^2}\delta(\omega-\omega')\slabel{eq:commEbar}\\
    &&\times \, \imag{G_{ij}(\rr,\rr',\omega)}.\nonumber 
\end{subeqnarray}
The above relations are further discussed in Appendix~\ref{app:rotatingFrame} and, from this point onward, operators will be considered exclusively in the rotating frame without explicitly noting the fact, e.g., "rotating-frame amplitude operators" will simply be referred to as "amplitude operators" and so on.

We find the equations of motion governing the evolution of the Cartesian components of $\vec{\opg}^{(+)}(\rr,\omega,t)$ from Eq.~\eqref{eq:rotFrameEquation} (the full derivation is given in Appendix~\ref{app:eqsOfMotionBar}):

\begin{eqnarray}\label{eq:gEqs}
\pdt\opg_i^{(+)}(\rr,\omega,t)&=&\int \dd{\rr'}\int \dd{\omega'} \nonumber\\
&&\times F_{ij}(\rr,\omega;\rr',\omega';t)\opg_j^{(-)}(\rr',\omega',t),\qquad
\end{eqnarray}
with the corresponding equation for $\pdt \opg_i^{(-)}(\rr,\omega,t)$ obtained by simply taking the adjoint of Eq.~\eqref{eq:gEqs}. Here we also defined: 
\begin{eqnarray}\label{eq:Fdefinition}
    F_{ij}(\rr,\omega;\rr',\omega';t)&=&\frac{2i}{\pi}\frac{\omega^2}{c^2}\imag{G_{il}(\rr,\rr',\omega)} \nonumber\\
    &&\times\pump{k}(\rr',t)  \chit_{klj}(\rr')\exp{i(\omega+\omega')t}.\qquad
\end{eqnarray}
Before proceeding with our derivation, we note that the equations of motion in Eq.(11) are derived under the condition of Kleinman’s symmetry (assuming a non-dispersive nonlinear susceptibility), where the $\chit_{ijk}$ tensor is symmetric under the permutation of its indices. Although this assumption is reasonable in many systems \cite{boyd2020nonlinear}, it is not necessary for the development of our formalism, and only results in simpler expressions. The case of Kleinman’s symmetry not being applicable is discussed in Appendix~\ref{app:eqsOfMotionBar}.

\subsection{Input-output relations}
The obtained equations of motion Eqs.~\eqref{eq:gEqs} are \textit{linear} in terms of the amplitude operators $\opg_{i}^{(\pm)}(\rr,\omega,t)$. Along with the fact that the $\opg_{i}^{(\pm)}(\rr,\omega,t)$ are themselves \textit{canonical}, i.e., their commutator Eq.~\eqref{eq:commEbar} is scalar, this enables us to claim that, at an arbitrary time $t$, $\opg_{i}^{(\pm)}(\rr,\omega,t)$ can be expressed as a linear combination of $\opg_{i}^{(\pm)}(\rr,\omega,t\rightarrow-\infty)$, i.e., the free-field amplitude operators \cite{serafini2017quantum,christ2013theory,wasilewski2006pulsed}.
Thus, we can postulate an ansatz for the general solution of Eqs.~\eqref{eq:gEqs} in which the time-dependent (output) operators are expressed in terms of the free-field (input) operators via an input-output (IO-) relation. Additionally, due to the linear relation between the amplitude operators $\opg_{i}^{(\pm)}(\rr,\omega,t)$ and creation/annihilation operators $\bar{\vec{f}}^{(\dagger)}(\rr,\omega,t)$, as can be seen from Eq.~\eqref{eq:gDef}, either the free-field amplitude operators or the free-field creation/annihilation operators can be used to formulate the IO-relation. For the purpose of this work, we use the creation/annihilation operators $\bar{\vec{f}}^{(\dagger)}(\rr',\omega,t\rightarrow-\infty)$, which we label $\bar{\vec{f}}^{(\dagger)}(\rr',\omega)$ for compactness. 
We note that a completely equivalent formalism is obtained if the amplitude operators $\vec{\opg}^{(\pm)}(\rr,\omega,t\rightarrow-\infty)$ are used instead and the main reason we opted for $\bar{\vec{f}}^{(\dagger)}(\rr',\omega)$ was to have more streamlined final expressions due to their simpler commutation relation in Eq.~\eqref{eq:commFbar}.
With the above discussion in mind, we assume the following ansatz for the IO-relation:
\begin{widetext}
\begin{equation}\label{eq:gIO}
    \opg_i^{(+)}(\rr,\omega,t)=\int\dd{\xx}\int\dd{\nu}\left(\mathcal{B}_{ij}(\rr,\omega;\xx,\nu;t)\bar{f}_j(\xx,\nu)+\conj{\mathcal{A}}_{ij}(\rr,\omega;\xx,\nu;t)\bar{f^\dagger}_j(\xx,\nu)\right).
\end{equation}
\end{widetext}
In the above expression, $\mathcal{A}_{ij}(\rr,\omega;\xx,\nu;t)$ and $\mathcal{B}_{ij}(\rr,\omega;\xx,\nu;t)$ are the \textit{input-output (IO)-coefficients}. In Eq.~\eqref{eq:gIO}, $\xx \text{ and } \nu$ are spatial and frequency variables, which enumerate the free-field field-matter local excitations.

The distribution of the contributing excitations is determined by the IO-coefficients, which are, in turn, determined by the properties of the system under consideration. A similar IO-relation is often derived in existing works on high-gain SPDC in lossless systems, where the output normal-mode photon creation/annihilation operators are expressed as linear combinations of normal-mode photon operators at the input, and is associated with frequency mixing caused by the nonlinear interaction \cite{wasilewski2006pulsed,christ2013theory}. In the present context, Eq.~\eqref{eq:gIO} does indeed represent the output as a linear combination of excitation of different frequencies, manifested by the frequency integral over $\nu$. However, it also shows the output to be a linear combination of excitations at different spatial positions, manifested by the spatial integral over $\xx$. This is a property inherent to the GF quantisation method and is a consequence of the local nature of the fundamental bosonic modes, which itself enables the treatment of arbitrary open optical systems. In the remainder of the paper, to make expressions more compact, we will combine the variables $\xx \text{ and } \nu$ into the vector $\XX=\left(\nu,\xx\right)$ whenever possible and replace $\int\dd{\xx}\int\dd{\nu}$ with $\int\dd{\XX}$.

Here, we again emphasise that the linearity of Eqs.~\eqref{eq:gEqs} in terms of the field operators, necessary for establishing the IO-relation Eq.~\eqref{eq:gIO}, is a direct consequence of the undepleted pump approximation and treating the pump field as a classical pulse (as shown in Appendix~\ref{app:eqsOfMotionBar}).

The input-output relation \eqref{eq:gIO} allows us to reconstruct the field operators at any time $t$, if the IO-coefficients $\mathcal{A}_{ij}(\rr,\omega,\XX;t)$ and $\mathcal{B}_{ij}(\rr,\omega,\XX;t)$ have been obtained. 
Moreover, Eq.~\eqref{eq:gIO} can be used to express field-dependent quantities directly in terms of the IO-coefficients. To illustrate this, we consider the first-order field correlation function $g_{ij}^{(1)}(\rr,\rr';t,t')=\expval{\ne_i(\rr,t)\pe_j(\rr',t')}$ as an example. Here, the expectation value is to be taken in the initial state of the system, since we are working in the Heisenberg picture. To write $g_{ij}^{(1)}(\rr,\rr';t,t')$ in terms of the IO-coefficients, we expand the field operators using Eq.~\eqref{eq:EwithSlowOps} and then replace the amplitude operators using the IO-relation Eq.~\eqref{eq:gIO}. If we assume the initial state to be the vacuum, we obtain (see Appendix~\ref{app:correlationF} for details):
\begin{eqnarray}\label{eq:firstOrderCorrel}
g_{ij}^{(1)}(\rr,\rr';t,t')&=& \iint \dd{\omega} \dd{\omega'}\exp{i\omega t-i\omega' t'}\int \dd{\XX} \nonumber \\ 
&&\times \mathcal{A}_{ik}(\rr,\omega,\XX;t)\mathcal{A}^*_{jk}(\rr',\omega',\XX;t').
\end{eqnarray}
If the system is initially not in the vacuum state, e.g., in the case of a seeded nonlinear process, $g_{ij}^{(1)}(\rr,\rr';t,t')$ (and any higher-order correlation function) can still be expressed in terms of the IO-coefficients, however, a non-vacuum initial state will also result in a more complicated expression, more details on this are given in Appendix~\ref{app:correlationF}.
To obtain numerical values of $g_{ij}^{(1)}(\rr,\rr';t,t')$, one need only find the IO-coefficients at times $t$ and $t'$ and insert them into Eq.~\eqref{eq:firstOrderCorrel}.\\

\subsection{Comparison with existing methods}
To make the benefits of the formalism discussed in this paper more apparent, before continuing, we will briefly review the differences and similarities between our approach and existing formalisms for treating high-gain SPDC. 
The IO-relation Eq.~\eqref{eq:gIO} is conceptually identical to Bogoliubov transformations used in Refs.~\cite{wasilewski2006pulsed,christ2013theory}, as well as the ansatz used in Ref.~\cite{chekhova2020BSV-increasingBrightness} – all of the aforementioned relations, including Eq.~\eqref{eq:gIO}, relate the output field operators to the creation and annihilation operators for the modes of the input field in a linear fashion. However, the key difference between these previous works and ours is the type of quantisation used for the field operators, which results in different definition of the modes of the quantised field, which in our case allows for direct inclusion of arbitrary amounts of linear losses. In Refs.~\cite{wasilewski2006pulsed,christ2013theory,chekhova2020BSV-increasingBrightness}, where loss/dissipation is completely neglected, one can use a quantisation scheme where the creation and annihilation operators correspond to the normal modes of the photonic system, which are power-orthogonal and spatially non-local. This results in output fields being expressed as linear combinations of normal-mode photon operators (distinguished either by frequency or wavevector). In the present context, with the GF quantisation method, the field operator is expressed in terms of local bosonic excitations of the field and matter, sometimes also called polaritonic modes \cite{knoll2003qed}. Such a description directly includes all the effects coming from the linear light-matter interactions in the photonic system (embedded in the GF), with potential loss channels. Some forms of linear losses can also be treated in the case of normal photonic modes, as has been done in Refs.~\cite{quesada2022beyond,triginer2020twin-beam,quesada-sipe2020twinBeam-waveguides},  through coupling the normal modes to weakly lossy channels. However, such treatment can only account for weak scattering and absorption losses, as it relies on assuming lossless/power-orthogonal normal modes for the underlying photonic system that are only weakly perturbed by the presence of the loss channels. The GF quantisation method can in turn also account for cases where loss is very strong and/or is inherently shaping the modal properties of the system, such as in strongly leaky systems (e.g. Mie nanoresonators \cite{weissflog2023nanoresonators}) or strongly lossy systems (e.g. plasmonic systems \cite{andrey2016prl}). This generality of our method comes at the cost of increased computational efforts, coming from the local nature of the creation and annihilation operators in the GF method, which requires carrying around the spatial degree of freedom in all the calculations.
Another notable difference between our formalism and existing ones is the absence of a weak-dispersion approximation, which is commonly used to simplify the equations of motion for the field operators by assuming a linear dispersion relation (between the wave-vector and frequency of the modes) and using the assumption to replace the time degree of freedom by the spatial one \cite{quesada2022beyond,wasilewski2006pulsed,christ2013theory,chekhova2020BSV-increasingBrightness,triginer2020twin-beam,quesada-sipe2020twinBeam-waveguides}. In certain cases, higher-order dispersion can also be included through additional terms in the equations of motion, corresponding to terms in a power-series expansion of the wave-vector of the photonic modes of the system \cite{quesada2022beyond,triginer2020twin-beam,quesada-sipe2020twinBeam-waveguides}. In order to make our formalism applicable to general nanostructured and open systems, with potentially highly complex dispersion relations for the involved photonic modes, we do not make any assumptions regarding the dispersion relation and solve our equations of motion temporally. This does come at the cost of increased computational effort when implementing the formalism numerically, but we note that the prescriptions for simplifying the resulting equations in cases where the weak-dispersion approximation is valid could also be applied to our formalism to obtain similar computational benefits.
Finally, we emphasize that generalizing the existing photon-based formalisms for inclusion of weak losses is usually done on a per-system basis, with focus on a specific system geometry (e.g. a waveguide or a ring resonator). In contrast, the formalism presented in our work offers a unified approach, applicable to a wide variety of open photonic systems, and is formulated in a way that is compatible with existing numerical methods to facilitate its usefulness.

\subsection{Coupled-mode equations for the IO-coefficients}
In order to calculate the IO-coefficients at a certain time, we find the coupled differential equations governing their time evolution by inserting Eq.~\eqref{eq:gIO} into the Heisenberg equations of motion Eq.~\eqref{eq:gEqs} as an ansatz to obtain:
\begin{subequations}
\label{eq:ABeqs}
\begin{eqnarray}
\pdt  \mathcal{A}_{ij}(\rr,\omega,\XX;t)&=&\int \dd{\rr'}\int\dd{\omega'}\conj{F}_{ik}(\rr,\omega;\rr',\omega';t)  \nonumber\\
&&\times\mathcal{B}_{kj}(\rr',\omega',\XX;t),\qquad \label{ABeqs:1}\\
\pdt \mathcal{B}_{ij}(\rr,\omega,\XX;t)&=&\int \dd{\rr'}\int\dd{\omega'}F_{ik}(\rr,\omega;\rr',\omega';t) \nonumber\\
&&\times \mathcal{A}_{kj}(\rr',\omega',\XX;t).\qquad\label{ABeqs:2}
\end{eqnarray}
\end{subequations}

The boundary conditions for the IO-coefficients are derived by equating Eq.~\eqref{eq:gIO} with Eq.~\eqref{eq:gDef} and setting $t\rightarrow-\infty$ on both sides:
\begin{subequations}
\label{eq:ABinit}
\begin{eqnarray}
    \mathcal{A}_{ij}(\rr,\omega;\XX;t\rightarrow-\infty)&=&0,\\
    \mathcal{B}_{ij}(\rr,\omega;\XX;t\rightarrow-\infty)&=& i\mathcal{K}\,\frac{\nu^2}{c^2}\sqrt{\epspp(\xx,\nu)} \nonumber \\
   &&\times G_{ij}(\rr,\xx,\nu)\delta(\omega-\nu).\quad \label{ABinit:2} 
\end{eqnarray}
\end{subequations}
In most practical cases, Eqs.~\eqref{eq:ABeqs} do not have analytical solutions and have to be solved numerically. To that end, we can rewrite them in a form more appropriate for implementation, which has the added benefit of more clearly distinguishing the different contributions to the output field in the IO-relation of Eq.~\eqref{eq:gIO}. We begin by decomposing the coefficient $\mathcal{B}_{ij}(\rr,\omega,\XX;t)$ in the following way:
\begin{equation}\label{eq:bDecomp}
    \mathcal{B}_{ij}(\rr,\omega,\XX;t)=b^{(0)}_{ij}(\rr,\omega,\XX)+B_{ij}(\rr,\omega,\XX;t)
\end{equation}
where $b^{(0)}_{ij}$ and $B_{ij}$ satisfy the following conditions:
\begin{eqnarray*}
b^{(0)}_{ij}(\rr,\omega,\XX)&\equiv& \mathcal{B}_{ij}(\rr,\omega,\XX;t\rightarrow-\infty)\\
\pdt B_{ij}(\rr,\omega,\XX;t)&=& \pdt \mathcal{B}_{ij}(\rr,\omega,\XX;t),\\
B_{ij}(\rr,\omega,\XX;t\rightarrow-\infty) &=&0.
\end{eqnarray*}
After inserting the decomposition Eq.~\eqref{eq:bDecomp} into Eq.~\eqref{eq:gIO} and relabeling $\mathcal{A}_{ij}\rightarrow A_{ij}$ for notational convenience, the IO-relation takes the form:
\begin{widetext}
\begin{equation}\label{eq:IOsourced}
    \opg_i^{(+)}(\rr,\omega,t)=\opg^{(0,+)}_i(\rr,\omega)+\int\dd{\XX}\left(B_{ij}(\rr,\omega,\XX;t)\bar{f}_j(\XX)+\conj{A}_{ij}(\rr,\omega,\XX;t)\bar{f}^\dagger_j(\XX)\right),
\end{equation}
\end{widetext}
where $\opg^{(0,+)}_i(\rr,\omega)=\int\dd{\XX}b^{(0)}_{ij}(\rr,\omega,\XX)\bar{f}_j(\XX)$ is exactly the free-field (input) amplitude operator of frequency $\omega$, since $\vec{\opg}^{(+)}(\rr,\omega,t\rightarrow-\infty)=\vec{\opg}^{(0,+)}(\rr,\omega)$. The remaining two terms in Eq.~\eqref{eq:IOsourced} quantify the changes that the free-field component of frequency $\omega$ undergoes due to the nonlinear interaction through the "new" set of coefficients -- $A_{ij}(\rr,\omega,\XX;t)$ and $B_{ij}(\rr,\omega,\XX;t)$.

The differential equations coupling the IO-coefficient $A_{ij}(\rr,\omega,\XX;t)$ and the newly introduced $B_{ij}(\rr,\omega,\XX;t)$ are derived by inserting the decomposition Eq.~\eqref{eq:bDecomp} into Eqs.~\eqref{eq:ABeqs}:\\
\begin{subequations}
\label{eq:sourceEqs}
\begin{eqnarray}
\pdt  A_{ij}(\rr,\omega,\XX;t)&=& S^{(0)}_{ij}(\rr,\omega,\XX;t) \nonumber \\
&&+\int \dd{\rr'}\int\dd{\omega'}\conj{F}_{ik}(\rr,\omega;\rr',\omega';t)\nonumber\\
&&\times B_{kj}(\rr',\omega',\XX;t),\label{sourceEq:1} \\
\pdt B_{ij}(\rr,\omega,\XX;t)&=&\int \dd{\rr'}\int\dd{\omega'} F_{ik}(\rr,\omega;\rr',\omega';t)\nonumber \\
&&\times A_{kj}(\rr',\omega',\XX;t).\label{sourceEq:2}
\end{eqnarray}
\end{subequations}
The coupled quantities in the above system have the initial values: $A_{ij}(\rr,\omega,\XX;t\rightarrow-\infty)=B_{ij}(\rr,\omega,\XX;t\rightarrow-\infty)=0$ and we defined $S^{(0)}_{ij}(\rr,\omega,\XX;t)=\int \dd{\rr'}\int\dd{\omega'}\conj{F}_{ik}(\rr,\omega;\rr',\omega';t)b^{(0)}_{kj}(\rr',\omega',\XX)$, which acts as a source term for equation \eqref{sourceEq:1}. After taking the frequency integral, it will have the following form:
\begin{eqnarray}\label{eq:sourceDef}
    S^{(0)}_{ij}(\rr,\omega,\XX;t)&=&\frac{2\mathcal{K}}{\pi} \frac{\omega^2 \nu^2}{c^4} \sqrt{\epspp(\xx,\nu)}\exp{-i(\omega+\nu)t}\nonumber \\
    &&\times\int \dd{\rr'}\chit_{klm}(\rr')\pumpn{k}(\rr',t)\nonumber \\
    &&\times\imag{G_{il}(\rr,\rr',\omega)}G_{mj}(\rr',\xx,\nu).\quad
\end{eqnarray}

The systems of coupled equations \eqref{eq:ABeqs} and \eqref{eq:sourceEqs}, along with their respective boundary conditions are equivalent formulations of our theoretical formalism and are the first part of the main results presented in this work. They are valid in arbitrary dispersive and open nanostructured systems, such as photonic crystals, nanoresonators, metasurfaces and waveguides, where the complex spatial and dispersive properties inherent to such devices can all be accounted for in the electromagnetic GF, as well as arbitrary loss, e.g. radiation leakage, material absorption, scattering losses etc.

To summarise the main results derived in this section: in order to find the output electric field operator of a high-gain SPDC system, our proposed formalism consists of solving the coupled equations \eqref{eq:sourceEqs}, which are defined by the system's classical Green's function $G_{ij}(\rr,\rr',\omega)$, linear susceptibility $\varepsilon(\rr,\omega)$ and nonlinear susceptibility $\chit_{ijk}(\rr)$. These properties, along with the pump field  $\vec{E}_P^{(\pm)}(\rr,t)$ are used to define the integration kernel $F_{ik}(\rr,\omega;\rr',\omega';t)$ and source term $S^{(0)}_{ij}(\rr,\omega,\XX;t)$, given by Eqs.~\eqref{eq:Fdefinition} and \eqref{eq:sourceDef}, respectively. Solving the equations provides the IO-coefficients, which are then inserted into the IO-relation Eq.~\eqref{eq:IOsourced} to obtain the output amplitude operators $\vec{\opg}^{(\pm)}(\rr,\omega,t)$. These, in turn enable us to reconstruct the full electric field operator as shown in Eq.~\eqref{eq:EwithSlowOps}.
The IO-coefficients can also be used to directly calculate scalar quantities related to the field, such as correlation functions, as was given in Eq.~\eqref{eq:firstOrderCorrel}.

\subsection{Frequency domain formulation}
The formalism developed in the previous section enables straightforward calculation of output fields and field-related quantities and is valid under very general considerations. However, the fields (and thus, all field-dependent quantities) are obtained in the \textit{time domain} and the calculation of frequency-domain fields and quantities, e.g., the spectra of output photons, is inefficient to calculate numerically from the obtained time-domain quantities. This can be seen by examining, for example, the expression for the single-photon spectrum of the electric field calculated using the IO-coefficients used in Eq.~\eqref{eq:IOsourced}.

The spectrum of a quantised electric field, measured at a time $t$ and position $\rr_0$ and summed over all field polarisations is given by \cite{cresser1983spectrum}:
\begin{eqnarray}\label{eq:prob}
	\sigma(\rr_0,\omega_0,t)&=&
	\iint\limits^{\quad t}_{-\infty}\dd{t'}\dd{t''}\exp{-i\omega_0(t'-t'')} \nonumber\\
	&&\times\expval{\vec{\ne}(\rr_0,t')\cdot\vec{\pe}(\rr_0,t'')}.\qquad
\end{eqnarray}
where $\omega_0$ is the frequency at which the spectrum is evaluated.

To obtain an expression for $\sigma(\rr_0,\omega_0,t)$ in terms of the IO-coefficients $A_{ij}(\rr,\omega,\XX;t)$ and $B_{ij}(\rr,\omega,\XX;t)$, we use the first-order correlation function Eq.~\eqref{eq:firstOrderCorrel} (remembering that $\mathcal{A}_{ij}\equiv A_{ij}$) and insert it into Eq.~\eqref{eq:prob}:
\begin{eqnarray}\label{eq:probWithA}
\sigma(r_0,\omega_0,t)&=&\iint\limits^{\quad t}_{-\infty}\dd{t'}\dd{t''}\iint\dd{\omega}\dd{\omega'}\int\dd{\XX} \nonumber \\
&&\times A_{ij}(\rr_0,\omega,\XX;t')\conj{A}_{ij}(\rr_0,\omega',\XX;t'')\nonumber\\
&&\times \exp{-i(\omega_0-\omega)t'}\exp{i(\omega_0-\omega')t''}. 
\end{eqnarray}
Although  Eq.~\eqref{eq:probWithA} theoretically allows one to calculate the spectrum, it can be impractical in a realistic case, when the IO-coefficients have to be obtained numerically. Namely, in order to perform the $t'/t'' \text{ and } \omega/\omega'$ integrals with sufficient numerical accuracy, one requires the values of the coefficient $A_{ij}(\rr,\omega,\XX;t)$ at many narrowly spaced times $t$ in addition to the other variables $\omega$ and $\XX$, which dramatically increases the computational resources required for such a calculation.

The above limitation is a consequence of the individual frequency amplitudes $\vec{\opg}^{(\pm)}(\rr,\omega,t)$ not being equivalent to the spectral components of the field that are detected in a spectrally resolved measurement in the presence of field sources, such as during a nonlinear field interaction \cite{eberly1977spectrum,cresser1983spectrum}.

To overcome these difficulties and further expand the applicability of our formalism, we develop a frequency-domain formulation, which can be applied in the calculation of various spectral quantities, while still rigorously including the effects of arbitrary loss and dispersion. It focuses on the "spectrally relevant" parts of the field, i.e. special field operators which directly contribute to the spectrum. These are defined as \textit{time-frequency} transforms of the time-domain fields \cite{eberly1977spectrum} and, as will be shown in the next sections, evolve in time in a way quite similar (although not identical) to $\vec{\opg}^{(\pm)}(\rr,\omega,t)$. They also have their own IO-relation, which enables us to use them as fundamental field variables when determining the time evolution of frequency-domain quantities.

We begin by noting that the field spectrum Eq.~\eqref{eq:prob} can also be written as:
\begin{equation}\label{eq:compactProb}
    \sigma(\rr_0,\omega_0,t)=\expval{\vec{\tilde{E}}^{(-)}(\rr_0,\omega_0,t)\vec{\tilde{E}}^{(+)}(\rr_0,\omega_0,t)},
\end{equation}
where the operators $\vec{\tilde{E}}^{(\pm)}(\rr,\omega,t)$ have been defined as:
\begin{equation}\label{eq:filteredE}
    \vec{\tilde{E}}^{(\pm)}(\rr,\omega,t)=\int_{-\infty}^t\dd{t'}\exp{\pm i \omega t'}\vec{\pme}(\rr,t').
\end{equation}
 We will refer to them as \textit{filtered} field operators, due to the definition Eq.~\eqref{eq:filteredE} being reminiscent of the action of a causal, ideally monochromatic filter on the time-domain field. In the absence of an interaction, they reduce to the free-field amplitude operators $\vec{\opg}^{(0,\pm)}(\rr,\omega)$ as $t\rightarrow\infty$. For finite $t$ and/or with sources (i.e. an interaction) present, their interpretation is more involved and is discussed in \cite{eberly1977spectrum,cresser1983spectrum}.

The expression for the spectrum in terms of the filtered field Eq.~\eqref{eq:compactProb} is immediately more appealing than Eq.~\eqref{eq:prob}, as it involves the expectation value of operators at a time $t$, when the spectrum itself is evaluated. Thus, the need for knowing the fields (or, equivalently, the IO-coefficients) at all times prior to $t$ is eliminated, provided the filtered fields $\vec{\tilde{E}}^{(\pm)}(\rr,\omega,t)$ can be obtained without much additional computational complexity, which, as will be shown in the following, is indeed the case.

As can be noted from Eq.~\eqref{eq:filteredE}, the filtered field operators are a linear transform of the "regular" $\vec{\hat{E}}^{(\pm)}(\rr,t)$ and thus of $\vec{\opg}^{(\pm)}(\rr,\omega,t)$, as well. This enables us to formulate an analogous input-output relation for the filtered fields, To that end, we combine Eq.~\eqref{eq:IOsourced} and Eq.~\eqref{eq:EwithSlowOps} and insert the resulting field decomposition into Eq.~\eqref{eq:filteredE}. Thus we obtain:
\begin{widetext}
\begin{equation}\label{eq:filteredIO}
\tilde{E}^{(+)}_i(\rr,\omega,t)=\tilde{E}^{(0,+)}_i(\rr,\omega)+\int\dd{\XX}\left(\tilde{B}_{ij}(\rr,\omega,\XX;t)\bar{f}_j(\XX)+\conj{\tilde{A}}_{ij}(\rr,\omega,\XX;t)\bar{f}^\dagger_j(\XX)\right),
\end{equation}
\end{widetext}
where we defined:
\begin{subequations}\label{eq:filteredABdef}
    \begin{eqnarray}
    \tilde{E}^{(0,+)}_i(\rr,\omega)&=&\int_{-\infty}^t\dd{t'}\int\dd{\omega'} \nonumber \\
    &&\times\exp{i(\omega-\omega')t'}\bar{E}^{(0,+)}_i(\rr,\omega'),\label{filteredFF}\\
    \tilde{A}_{ij}(\rr,\omega,\XX;t)&=&\int_{-\infty}^t\dd{t'}\int\dd{\omega'} \nonumber\\
    &&\times \exp{-i(\omega-\omega')t'}A_{ij}(\rr,\omega',\XX;t'),\label{filteredABdef:1}\\
    \tilde{B}_{ij}(\rr,\omega,\XX;t)&=&\int_{-\infty}^t\dd{t'}\int\dd{\omega'} \nonumber\\
    &&\times\exp{i(\omega-\omega')t'}B_{ij}(\rr,\omega',\XX;t').\label{filteredABdef:2}
    \end{eqnarray}
\end{subequations}
Here, $\tilde{E}^{(0,+)}_i(\rr,\omega)$ is the filtered analogue of the free-field amplitudes $\bar{E}^{(0,+)}_i(\rr,\omega)$ and $\tilde{A}_{ij}(\rr,\omega,\XX;t)$ and $\tilde{B}_{ij}(\rr,\omega,\XX;t)$ are the IO-coefficients of the filtered field. For conciseness, we will refer to them as \textit{filtered IO-coefficients}. 

An expression for the spectrum in terms of the filtered coefficients can be found by grouping time and frequency integrals with their corresponding $A_{ij}$ coefficient in Eq.~\eqref{eq:probWithA} and recognising that the resulting integral quantities exactly match the definitions in Eqs.~\eqref{eq:filteredABdef}. The expression thus obtained is:
\begin{eqnarray}\label{eq:probWithAtilde}
\sigma(\rr_0,\omega_0,t)=
\int\dd{\XX} \tilde{A}_{ij}(\rr_0,\omega_0,\XX;t)\conj{\tilde{A}}_{ij}(\rr_0,\omega_0,\XX;t).\qquad
\end{eqnarray}
As expected, we no longer require knowledge of the IO-coefficients at narrowly spaced times, unlike Eq.~\eqref{eq:probWithA}, - we only need the filtered coefficients at time $t$.
To find their values at a given time, we obtain a new set of coupled equations by combining the definitions Eqs.~\eqref{eq:filteredABdef} with the system Eq.~\eqref{eq:sourceEqs} (the full derivation can be found in Appendix~\ref{app:filteredEqs}):
\begin{subequations}
\label{eq:tildeABeqs}
\begin{eqnarray}
\pdt  \tilde{A}_{ij}(\rr,\omega,\XX;t)&=& \tilde{S}^{(0)}_{ij}(\rr,\omega,\XX;t) \nonumber\\
&&+\int \dd{\rr'}\int\dd{\omega'} \conj{\tilde{F}}_{ik}(\rr,\omega;\rr',\omega';t) \nonumber \\
&&\times\tilde{B}_{kj}(\rr',\omega',\XX;t), \label{tildeABeqs:1}\\
\pdt \tilde{B}_{ij}(\rr,\omega,\XX;t)&=&\int \dd{\rr'}\int\dd{\omega'} \tilde{F}_{ik}(\rr,\omega;\rr',\omega';t)  \nonumber\\
&&\times\tilde{A}_{kj}(\rr',\omega',\XX;t),\qquad\label{tildeABeqs:2}
\end{eqnarray}
\end{subequations}
where $\tilde{S}^{(0)}_{ij}(\rr,\omega,\XX;t)$ is derived from the source term $S^{(0)}_{ij}(\rr,\omega,\XX;t)$, present in Eq.~\eqref{sourceEq:1} and is defined as:
\begin{eqnarray}\label{eq:sourceDefTilde}
    \tilde{S}^{(0)}_{ij}(\rr,\omega,\XX;t)&=&2i\mathcal{K}\, \frac{\nu^2}{c^2} \sqrt{\epspp(\xx,\nu)}\exp{-i(\omega+\nu)t} \nonumber \\
    &&\times
    \int \dd{\rr'}\chit_{klm}(\rr')G_{mj}(\rr',\xx,\nu)\nonumber \\
    &&\times\int\dd{\omega_p}\frac{(\omega_p-\nu)^2}{c^2}\fpumpn{k}(\rr',\omega_p) \nonumber \\
    &&\times\conj{G}_{il}(\rr,\rr',\omega_p-\nu)\exp{i\omega_p t}.
\end{eqnarray}
Here, $\mathcal{E}_{P,k}^{(\pm)}(\rr,\omega_p)$ is the Fourier transform of the pump amplitude, defined by $\vec{E}_P^{(\pm)}(\rr,t)=\int\dd{\omega_p}\vec{\mathcal{E}}_P^{(\pm)}(\rr,\omega_p)\exp{\mp i\,\omega_p t}$.
The integral kernel $\tilde{F}_{ik}(\rr,\omega;\rr',\omega';t)$ which couples the two coefficients in Eq.~\eqref{eq:tildeABeqs} is given by:
\begin{eqnarray}
\tilde{F}_{ik}(\rr,\omega;\rr',\omega';t)&=& \frac{2i}{\pi}\int\dd{\bar{\omega}}\frac{\bar{\omega}^2}{c^2}\chit_{lmk}(\rr')\fpump{l}(\rr',\omega'+\bar{\omega}) \nonumber\\
&&\times\imag{G_{im}(\rr,\rr',\bar{\omega})}\exp{i(\omega-\bar{\omega})t},\label{eq:fDefTilde}
\end{eqnarray}
and the initial conditions for the filtered coefficients are easily inferred from the definitions \eqref{eq:filteredABdef}:
$$\tilde{A}_{ij}(\rr,\omega,\XX;t\rightarrow -\infty)=\tilde{B}_{ij}(\rr,\omega,\XX;t\rightarrow -\infty)\equiv0.$$

The coupled equations Eqs.~\eqref{eq:tildeABeqs} are the second part of our main results and can be summarised as follows: to find the output field in the frequency domain, one must solve the coupled equations \eqref{eq:tildeABeqs} to obtain the filtered IO-coefficients. As was the case in the time-domain formulation, the properties of the optical system and the pump field are embedded in the source term $\tilde{S}^{(0)}_{ij}(\rr,\omega,\XX;t)$, defined in Eq.~\eqref{eq:sourceDefTilde}, and the integration kernel $\tilde{F}_{ik}(\rr,\omega;\rr',\omega';t)$, given by Eq.~\eqref{eq:fDefTilde}. The filtered IO-coefficients can then be used to reconstruct the frequency domain field operators using the IO-relation Eq.~\eqref{eq:filteredIO}. Alternatively, the filtered IO-coefficients can also be used to directly evaluate frequency-domain expectation values, such as the field spectrum Eq.~\eqref{eq:probWithAtilde}, discussed in the beginning of this section, or field moments of the form $\expval{\tilde{E}^{(-)}_i(\rr,\omega,t)\tilde{E}^{(+)}_j(\rr,\omega',t)}$ and $\expval{\tilde{E}^{(+)}_i(\rr,\omega,t)\tilde{E}^{(+)}_j(\rr,\omega',t)}$. The field moments can be used to reconstruct the joint spectral amplitude (JSA) of the output state of the system \cite{quesada2022beyond}, which is extremely useful when investigating high-gain effects, e.g. squeezing \cite{sharapova2015schmidtModes}.

To verify the results of our frequency-domain approach, we find the low-gain analytical solutions to Eqs.~\eqref{eq:tildeABeqs} and insert them into Eq.~\eqref{eq:probWithAtilde} to obtain the low-gain signal photon spectrum. These calculations are shown in Appendix~\ref{app:perturbative} and the obtained expression for the spectrum, shown in Eq.~\eqref{eq:probPert}, matches the results obtained in previous works on low-gain SPDC that used the GF quantisation formalism \cite{andrey2016prl,saravi2017amspdc}.

\section{Integrated quantum spectroscopy with undetected photons in the high-gain regime}\label{sec:qsup}
Quantum spectroscopy with undetected photons (QSUP) is a spectroscopic technique which relies on frequency-entangled, non-degenerate photon pairs, where only one of the entangled photons (e.g., the idler) interacts with the measured sample, but the effects of the sample’s dispersion and/or absorption can be studied by detecting the other photon (e.g., the signal). This method can be used as a way to overcome the limited availability of high-sensitivity detectors in certain frequency regions, e.g., infrared or terahertz \cite{kalashnikov2016infrared,lee2019qsup,lemos2022quantum}.

In Ref.~\cite{kumar2020integrated}, a QSUP scheme was proposed, in which a material to be sensed is placed in the near-field of a waveguide SPDC source. In such a configuration, when the material has a spectrally localised absorption line around the frequency range of the idler photons, it also affects the spectral properties of the generated signal photons \cite{kumar2020integrated,saravi2017amspdc}. In the aforementioned work, the proposed scheme was investigated perturbatively, in the low-gain regime of SPDC. Using our formalism, we can now extend the investigation into the high-gain regime through numerical simulations of integrated QSUP at varying amounts of gain and show that the spectral sensitivity of the scheme improves at higher parametric gains. 

We note here that a similar improvement of sensitivity at high gain has already been observed in other schemes involving measurements with undetected photons, such as quantum imaging and optical coherence tomography systems based on nonlinear interferometers and induced coherence \cite{kolobov2017controlling,giese2017phase,machado2020oct}. In these types of schemes, a sample is placed between two coherently pumped high-gain SPDC sources, which interacts with the output idler arm of the first source after generation.
In contrast, the integrated QSUP scheme proposed and investigated in the low-gain regime in \cite{kumar2020integrated}, involves an analyte sample interacting with the nonlinear waveguide during the pair-production process, resulting in the spectroscopic information about the sample being imparted onto the spectrum of the produced photon pairs. The frequency domain formulation of our formalism is ideally suited for investigating this kind of scheme, as it enables us to study the spectral properties of the output photons at high parametric gains in the presence of significant, spectrally-varying loss, that is directly affecting the SPDC process at the generation stage. 

\begin{figure}[hbtp]
\includegraphics[width=\linewidth]{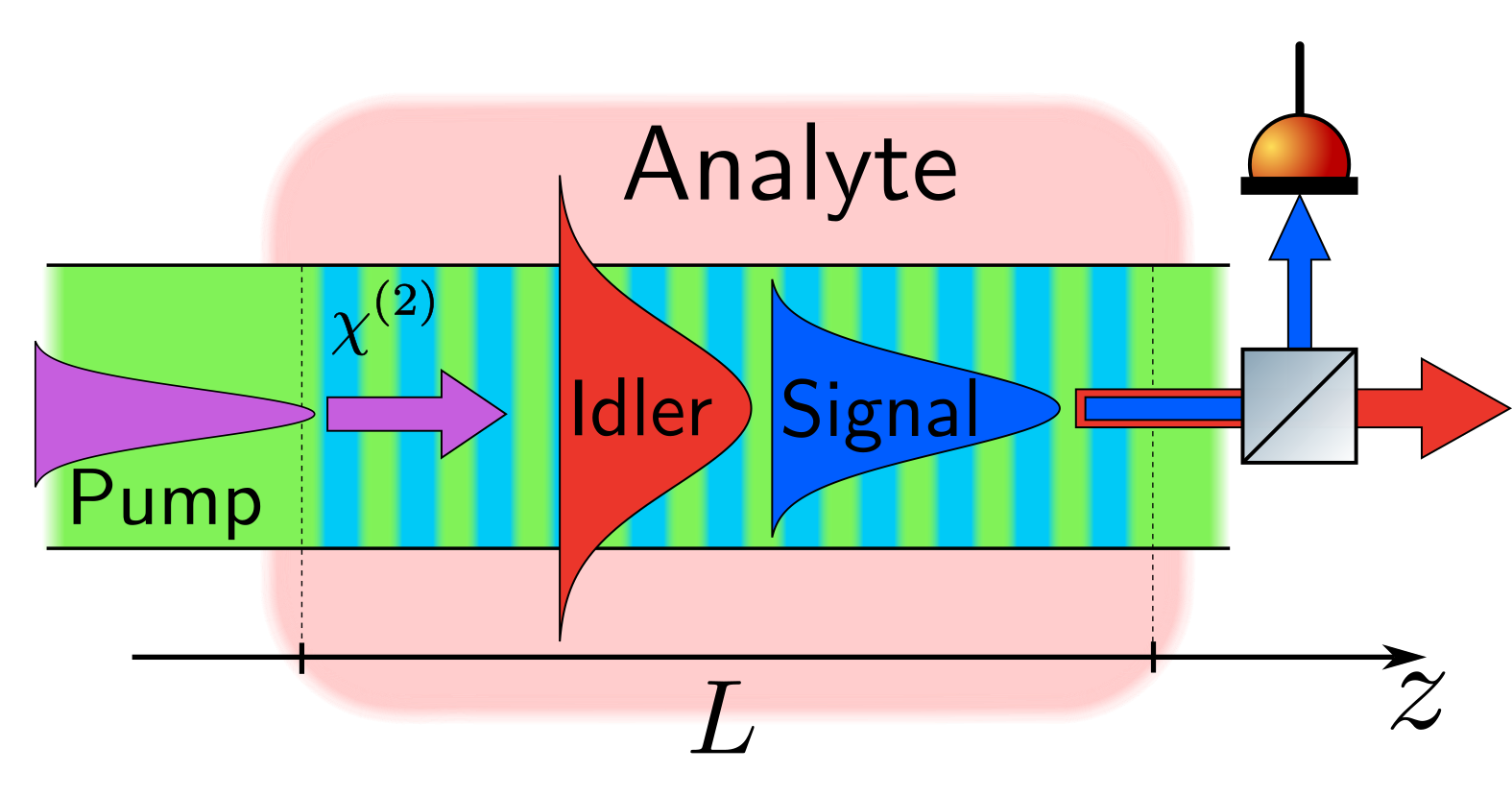}
\caption{\label{fig:schematic} Schematic representation of the integrated quantum spectroscopy process. A nonlinear waveguide SPDC source of length $L$ produces signal (blue) and idler (red) photons when excited by a pump pulse (purple). The idler mode, with a larger wavelength and consequently broader mode profile, interacts with the analyte (pale red) which has an absorption line in the frequency range of the idler photons. At the output of the waveguide, the signal and idler photons are separated by a beamsplitter which directs the signal photons into a spectrally-resolving detector while the idler photons remain undetected. The periodic light-blue-green region represent the periodically poled region of the nonlinear waveguide, allowing efficient phase-matching within the length $L$.}
\end{figure}

We consider the system shown schematically in Fig.~\ref{fig:schematic}: a periodically poled nonlinear waveguide, oriented along the $z$-axis, is excited by a pump pulse with a Gaussian temporal envelope. For simplicity, we neglect any polarisation or transverse dependence and assume that the waveguide is homogeneous, dispersive and lossy, characterised by the complex, frequency-dependent permittivity $\varepsilon(\omega)$, which includes both the effect of the waveguide plus an analyte that is interacting with the near-field of the waveguide modes.
Under the assumptions noted here, the GF of the waveguide has the analytical form \cite{gruner1996gf-quant}:
\begin{equation}\label{eq:1dGF}
    G(z,z',\omega)=\frac{i}{2\frac{\omega}{c}\sqrt{\varepsilon(\omega)}}\exp{i\frac{\omega}{c}\sqrt{\varepsilon(\omega)}\abs{z-z'}}.
\end{equation}
Since many relevant analytical properties of the GF are direct consequences of the Kramers-Kronig relations between the real and imaginary parts of the dielectric permittivity, absorption and dispersion of the model medium must be consistent with the aforementioned relations to ensure physically meaningful results \cite{gruner1996gf-quant}.
Accordingly, we use a Lorentz model for the permittivities of the waveguide dielectric and the analyte \cite{saleh-teich2019}. The permittivity of the combined waveguide and analyte system has the form:
\begin{equation}\label{eq:fullPermittivity}
\varepsilon(\omega)=1+\frac{\Omega_{Pl}^2}{(\Omega_0^2-\omega^2-i\Gamma \omega)}+\alpha\, \varepsilon_\mathrm{loss}(\omega)
\end{equation}
where $\Omega_{Pl}$ is the plasma frequency of the waveguide dielectric, while $\Omega_0$ and $\Gamma$ are  the frequency and width of the dielectric resonance, respectively. The term $\varepsilon_\mathrm{loss}(\omega)=\frac{\Omega_{Pl}^2}{(\omega_\mathrm{loss}^2-\omega^2-i\gamma \omega)}$, models the spectrally localised loss of the analyte, whose magnitude is modulated through the unitless factor $\alpha$. Here, $\omega_\mathrm{loss}$ and $\gamma$ are the frequency and width of the analyte resonance, respectively. The plasma frequency of the analyte is assumed to be the same as the waveguide for simplicity.

The factor $\alpha$ represents effective overlap of the evanescent tails of the guided mode at the idler frequency with the surrounding analyte. More specifically, we are assuming a scenario similar to Ref.~\cite{kumar2020integrated}, where the mode at the longer-wavelength idler has a much stronger overlap with the analyte, compared to the shorter-wavelength signal and pump. For a given waveguide geometry, the value of $\alpha$ can be determined by numerically finding the eigenmodes of the waveguide and calculating the effective permittivities of the involved modes \cite{kumar2020integrated}. In our simulations, the value was chosen to be $\alpha=3\times10^{-8}$, corresponding to only a small portion of the idler mode interacting with the analyte, which was nevertheless sufficient for our investigations. We note that in our model, to focus on the main physics, we are not considering the waveguide-geometry-dependent modal dispersions, which again, can be calculated through a rigorous eigenmode solver, and can be directly inserted into our model by substituting $\varepsilon(\omega)$ with a numerically calculated effective quantity.

For the incoming pump pulse, we use the frequency-domain form: $\mathcal{E}_P^{(+)}(z,\omega)=E_0\sqrt{\tau_p}\exp{-2\tau_p^2(\omega_{p_0}-\omega)^2}\exp{i k(\omega) z}$. Here, $\omega_{p0}$ is the pump central frequency, $\tau_p$ is the temporal width of the pulse, $k(\omega)=\frac{\omega}{c}\sqrt{\varepsilon(\omega)}$ and $E_0$ is a normalisation constant determined by the total energy of the pump pulse $U_0$, where $U_0\propto\abs{E_0}^2$.

\subsection{Simulation parameters}
All of the simulation parameters given here were normalised in the following manner: frequency quantities are expressed in units of $\omega_{p0}$, temporal quantities in units of $\frac{1}{\omega_{p0}}$ and spatial quantities in units of $\frac{2 \pi c}{\omega_{p0}}$ (the central wavelength of the pump in vacuum).
The nonlinear region of the waveguide is centred at the coordinate origin with the length $L=3\times10^{4}$ and is periodically poled with a spatial dependence $\chit(z)=\chit_m\cos{\Lambda z}$, where $\chit_m$ is the maximum absolute value of the nonlinear permittivity and $\Lambda$ is the poling period, chosen such that the central phase-matched frequencies of the signal and idler photons are $\omega_{s0}=0.7$ and $\omega_{i0}=0.3$, respectively. 
The resonance frequency of the analyte was set to be identical to the idler central phase-matching frequency, $\omega_\mathrm{loss}=\omega_{i0}=0.3$, to make the effects of the loss more prominent. The width of the loss spectrum was chosen to be $\gamma=2.5\times10^{-3}$.
The waveguide parameters in $\varepsilon(\omega)$ were set to be $\Omega_0=2.1$, $\Omega_{Pl}=0.25$ and $\Gamma=10^{-7}$, in order to satisfy the following conditions:
\begin{enumerate}[label=\roman*)]
    \item the resonance of the dielectric is far above the frequency region of interest and is narrow enough so the dielectric is effectively lossless for frequencies around and below $\omega_{p0}$; 
    \item the length of the nonlinear region, in combination with the refractive index of the dielectric, ensures a phase-matching bandwidth significantly wider than the width of the loss spectrum;
    \item the pump temporal width was chosen to be $\tau_p=2400$, resulting in a bandwidth sufficiently narrower than the loss spectrum to allow for observing the spectral correlation between the signal photon spectrum and the spectrum of the loss.
\end{enumerate}
The strength of the nonlinear interaction, which contributes to determining the parametric gain, is characterised by the product $\chit_m E_0\propto \chit_m \sqrt{U_0}$, which is varied to simulate SPDC at different values of parametric gain. In our case, this is done by adjusting the pump pulse energy $U_0$, while keeping the maximum nonlinear susceptibility $\chit_m$ constant. 

Finally, all of our simulations were performed for a sufficiently long time interval, in which the pump has had enough time to completely pass through the structure.

\subsection{Lossless medium}\label{sec:lossless}
We begin our investigation by studying the single-photon spectrum of SPDC at different values of parametric gain in the "lossless" case, i.e. without an analyte (the loss associated with the waveguide dielectric, although negligibly small in the frequency region of interest, is nevertheless accounted for in $\varepsilon(\omega)$). For different values of pump pulse energy $U_0$, we calculate the IO-coefficients by numerically solving Eqs.~\eqref{eq:tildeABeqs}, which are then used to evaluate the spectrum, as per Eq.~\eqref{eq:probWithAtilde}. 

\begin{figure}[htpb]
\includegraphics[width=\linewidth]{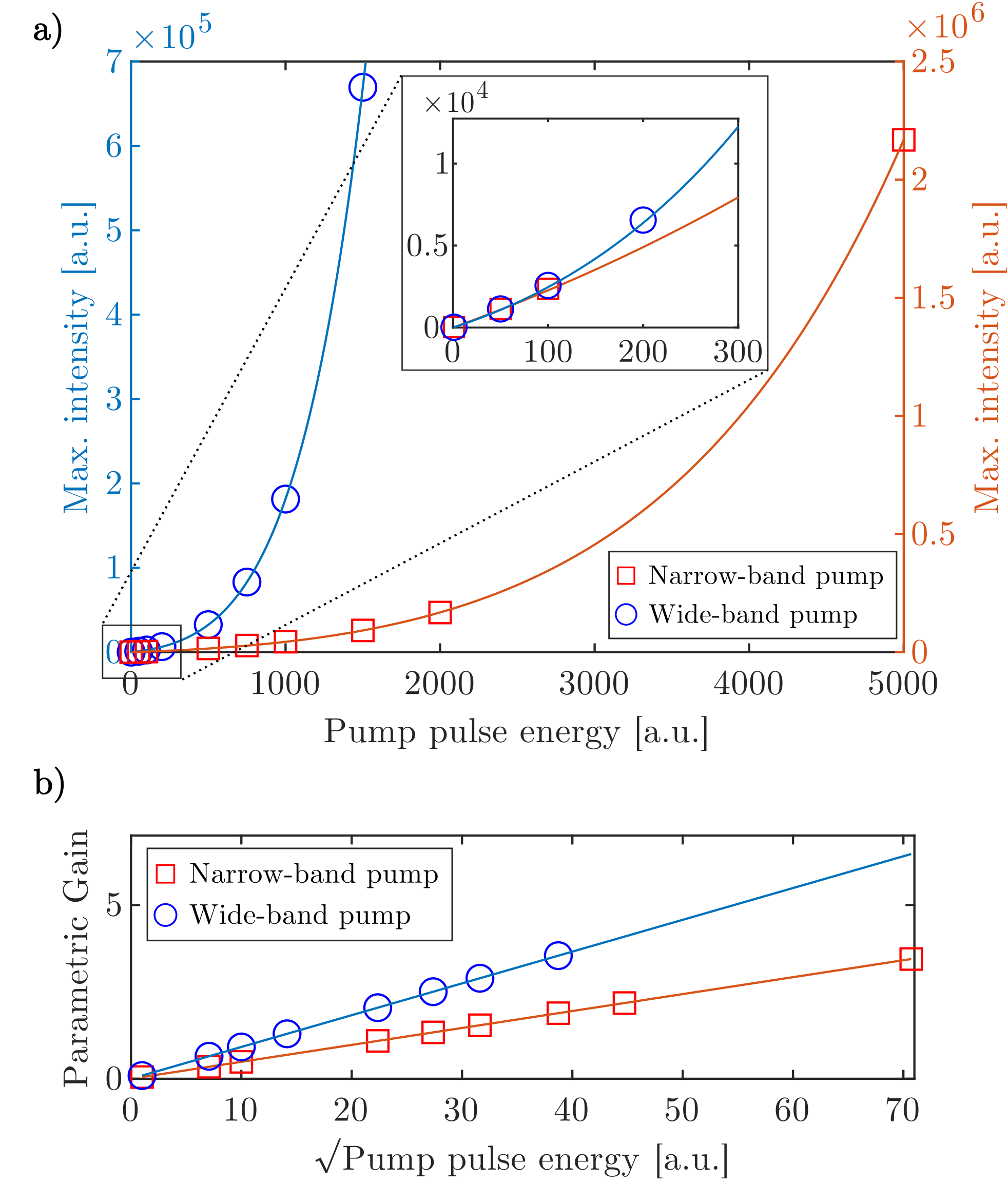}
\caption{\label{fig:gains}(\textbf{a}) The dependence of the spectral maximum at $\omega=0.7$ on the total pump energy for the narrow-band pump (red squares) with $\tau_p=2400$ and wide-band pump (blue circles) with $\tau_p=600$. The solid lines indicate the fitted $\sinh^2{(b\sqrt{U_0})}$ and the inset shows the intensity dependencies at very low pump pulse energies. In this regime, the behaviour of the spectrum becomes independent of the pump bandwidth. (\textbf{b}) The dependence of the observed parametric gain on the square root of the pump energy for the narrow- (red squares) and wide-band pump pulses (blue circles). In accordance with previous works, the wide-band pump results in higher parametric gain for a given pump pulse energy.}
\end{figure}

To extract the values of parametric gain associated with different pump intensities, we follow the prescription detailed in Ref.~\cite{chekhova2020BSV-increasingBrightness}: we plot the dependence of the maximal intensity of the single-photon spectrum $\mathcal{I}_m$ (at $\omega_s=\omega_{s0}=0.7$) as a function of the pump energy $U_0$ and fit it with the well-known dependence of the single-mode intensity in the case of a two-mode squeezer: $\mathcal{I}_m=a \sinh^2{(b\sqrt{U_0})}$, where $a$ and $b$ are fitting parameters. The parametric gain $\mathcal{G}$ for a particular value of pump energy is then defined as $\mathcal{G}=b\sqrt{U_0}$. 
In Fig.~\ref{fig:gains}a, we show the dependence of the spectrum maximum on pump pulse energy for pump pulses of two different bandwidths, defined by their temporal widths $\tau_p=2400$ and $\tau_p=600$, and observe that they both indeed follow the $\sinh^2$ law. The wider-bandwidth pulse, corresponding to $\tau_p=600$, was included in our simulations to make sure our formalism correctly predicts that a wider pump spectrum results in higher parametric gain for the same total pulse energy \cite{sharapova2015schmidtModes,chekhova2020BSV-increasingBrightness}. The inset of Fig.~\ref{fig:gains}a shows that the maximal intensity is approximately independent of the pump bandwidth at low pump energies, again, in accordance with previous observations \cite{sharapova2015schmidtModes,chekhova2020BSV-increasingBrightness}.
The dependence of the parametric gain on the pulse energy is shown in Fig.~\ref{fig:gains}b, where the higher amount of parametric gain obtained for wider pump bandwidths is evident.

\begin{figure}[hpb]

\includegraphics[width=0.95\linewidth]{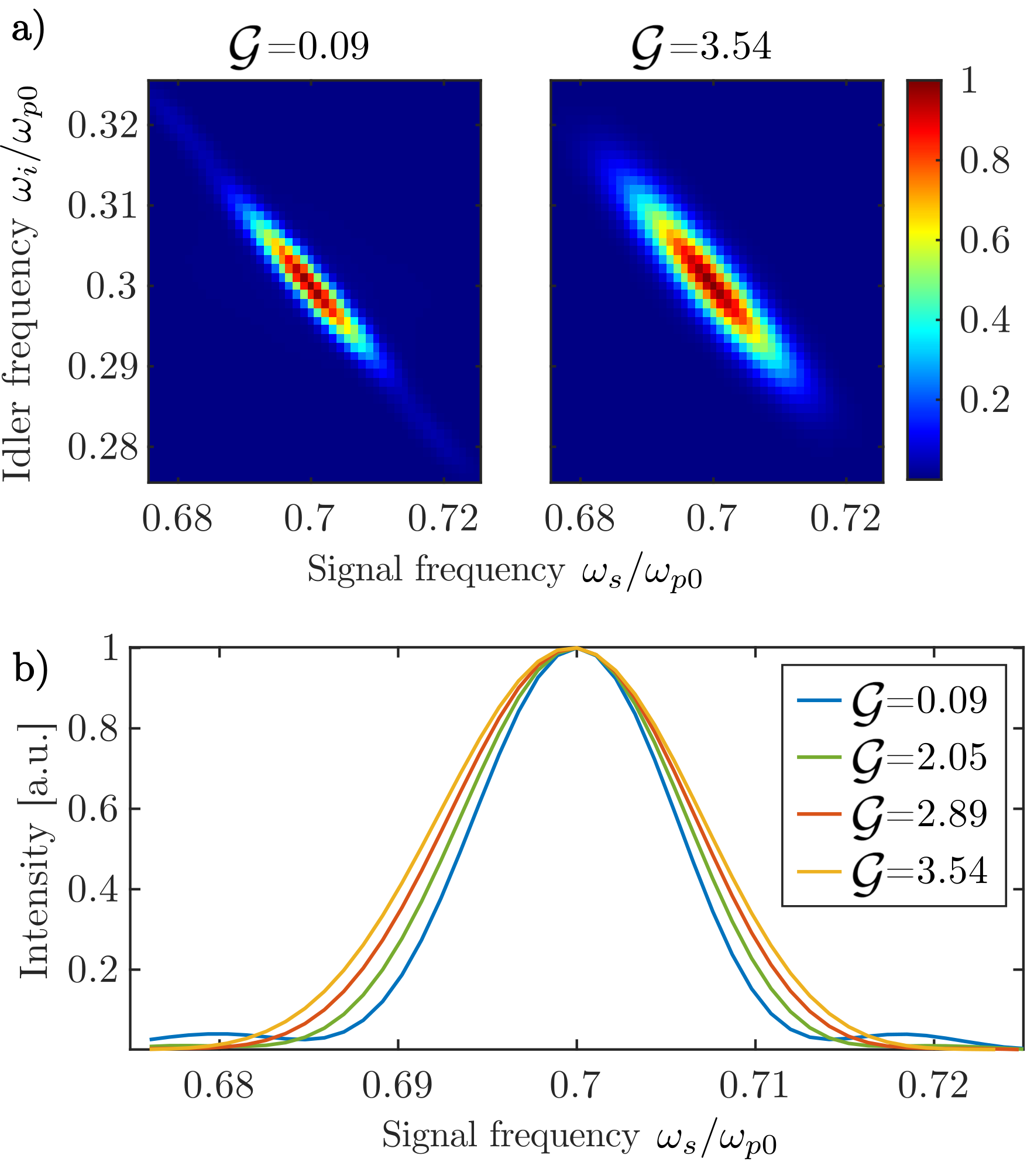}
\caption{\label{fig:moments}
(\textbf{a}) The normalised second-order field moment $N(\omega_i,\omega_s)$ for the wide-band pump ($\tau_p=600$)  at low parametric gain - $\mathcal{G}=0.0915$ and high parametric gain - $\mathcal{G}=3.54$. The output state at higher parametric gain exhibits less frequency correlations between the photons due to gain-induced broadening. (\textbf{b}) Normalised single-photon spectrum of the signal photons at the output (each spectrum normalised to its maximum) for the wide-band pump with $\tau_p=600$ at varying gain.}
\end{figure}

Aside from the behaviour of the spectrum amplitudes, our formalism also correctly predicts the broadening of the single- and two-photon spectra, occurring as gain is increased \cite{sharapova2015schmidtModes,chekhova2020BSV-increasingBrightness}. The broadening of the two-photon spectrum can be observed by studying the second-order moment of the output field $N(\omega,\omega')=\expval{\vec{\tilde{E}}^{(+)}(\rr_0,\omega,t)\cdot \vec{\tilde{E}}^{(+)}(\rr_0,\omega',t)}$, which is related to the JSA of the output photons \cite{quesada-sipe2020twinBeam-waveguides}. We can calculate the moment directly in terms of the IO-coefficients using the expression:
\begin{equation}
    N(\omega,\omega')=\int\dd{\XX}\tilde{B}_{ij}(\rr_0,\omega,\XX;t)\conj{\tilde{A}}_{ij}(\rr_0,\omega',\XX;t),
\end{equation}
which is obtained by expanding the filtered field operators using Eq.~\eqref{eq:filteredIO} and taking the expectation value.
$N(\omega_i,\omega_s)$, in the case of wide-band pump ($\tau_p=600$), is shown in Fig.~\ref{fig:moments}a for two values of parametric gain and we observe that higher gain indeed results in the moment becoming broader in frequency and, correspondingly, in output photons to exhibit less frequency correlations. Additionally, in Fig.~\ref{fig:moments}b, we show the normalised single photon spectrum at different values of gain for the wide-band pump. As gain is increased, the spectrum broadens, in accordance with previous results on high-gain lossless SPDC \cite{sharapova2015schmidtModes,chekhova2020BSV-increasingBrightness}.

\subsection{Lossy medium}\label{sec:lossy}
\begin{figure}[hptb]
\includegraphics[width=\linewidth]{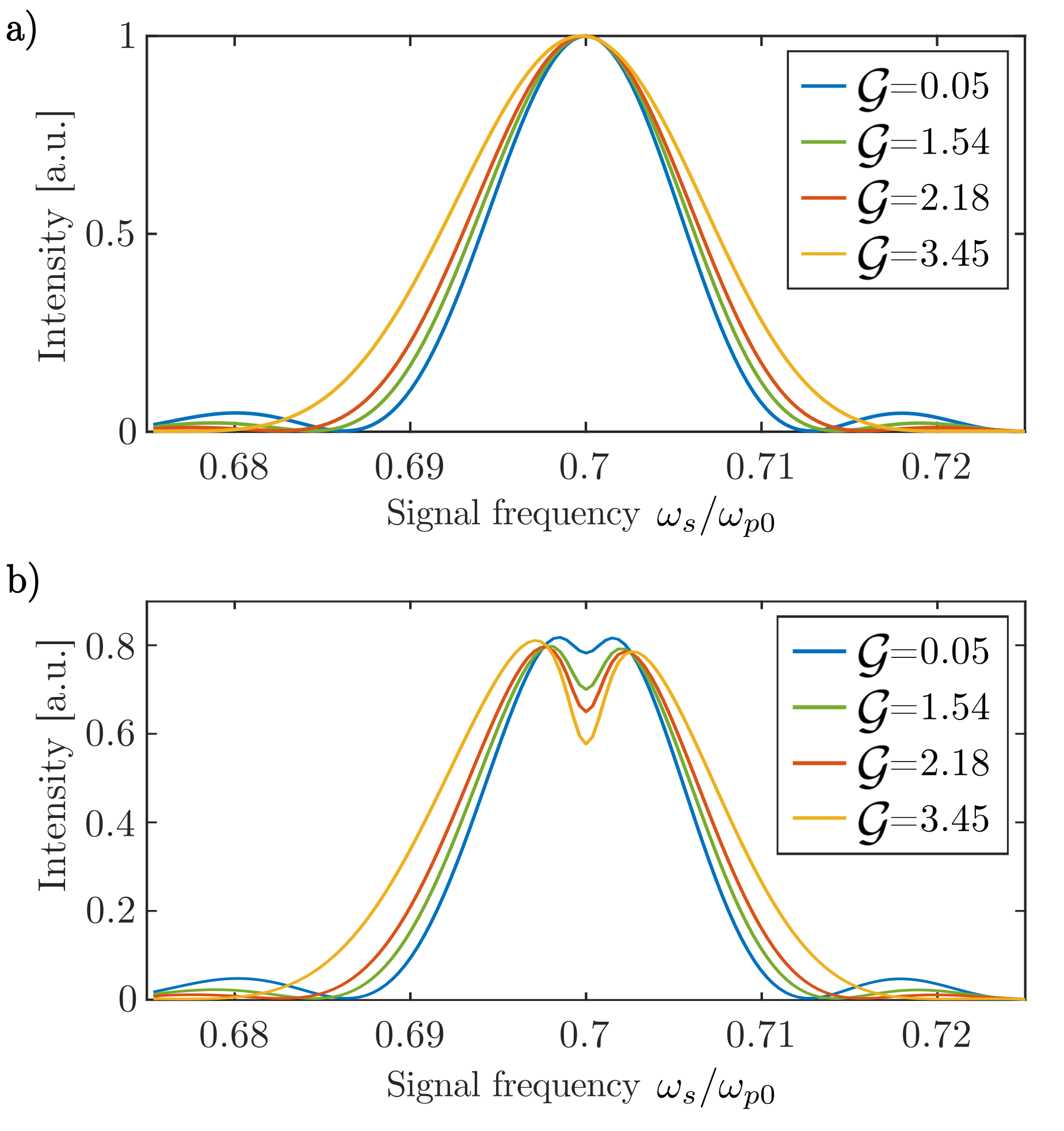}
\caption{\label{fig:lossySpectrum}Normalised single-photon spectrum of the signal photons for the narrow-band pump with $\tau_p=2400$ at varying amounts of parametric gain,  (\textbf{a}) without the analyte, and (\textbf{b}) with the analyte. The analyte loss is centred around $\omega_\mathrm{loss}=0.3$. In the lossy case, each spectrum is normalised to its lossless counterpart with the same gain, in order to better showcase the changes in the spectral shape of the dip.}
\end{figure}

To investigate the effects of increasing gain in QSUP, we introduce the analyte into the material permittivity $\varepsilon(\omega)$ and calculate signal photon spectra at the output.
In Fig.~\ref{fig:lossySpectrum}a we show the output spectra without the analyte and in Fig.~\ref{fig:lossySpectrum}b we show the spectra with the analyte present. The effect of the idler loss is immediately evident, as the spectra show a dip in the signal intensity, centred around the frequency $\omega^{(c)}=0.7$, which corresponds to the resonance frequency of the analyte $\omega_\mathrm{loss}$ through the conservation of energy $\omega_{p0}=\omega^{(c)}+\omega_\mathrm{loss}$. Both sets of spectra were obtained for the narrow-band pump with $\tau_p=2400$. 

\begin{figure}[htpb]
\includegraphics[width=\linewidth]{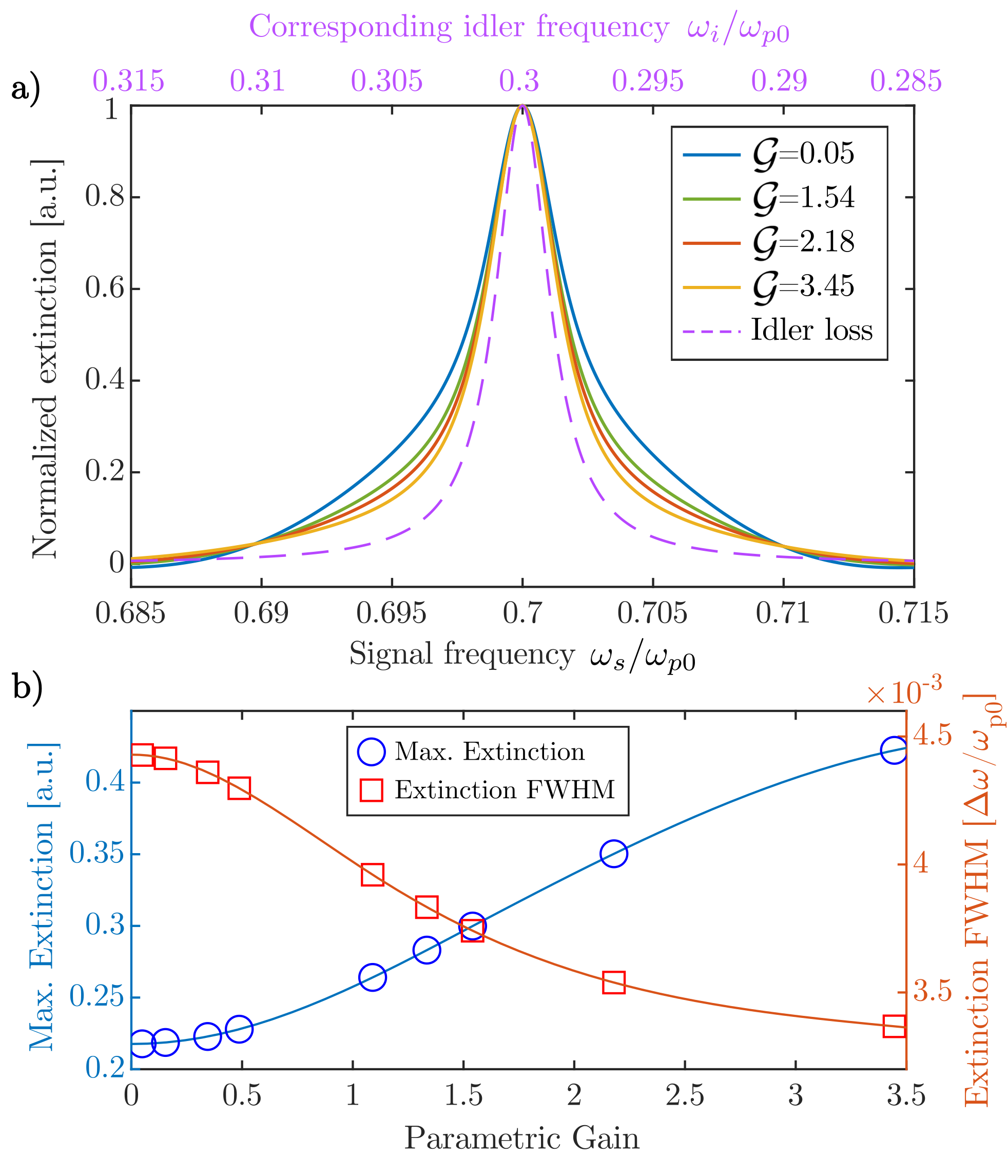}
\caption{\label{fig:extinction} (\textbf{a}) The extinction spectra of the signal photon at different values of parametric gain (solid lines) and the lineshape of the idler loss, namely $\imag{\varepsilon_\mathrm{loss}}$ (dashed line). The plots are all normalised to their own maximum amplitude to showcase the extinction spectrum becoming narrower as gain is increased. (\textbf{b}) The dependence of the extinction maximum (blue circles) and the FWHM of the extinction spectrum on parametric gain (red squares). The solid lines are interpolation curves to help illustrate the observed tendencies of the extinction spectrum.}
\end{figure}

The appearance of the spectral dip in the presence of idler loss is in accordance with the perturbative results of Ref.~\cite{kumar2020integrated}, where the perturbative calculations also suggested the spectral shape of the dip (i.e. its relative depth and width) to be independent of gain. However, using our non-perturbative formalism, we observe that the depth and spectral width of the dip does change as gain is increased. To quantify and investigate this behaviour we calculate the signal photon extinction spectrum (defined as the difference between the lossless and lossy spectra) for different values of gain. The extinction spectra and their properties at different values of gain are shown in Fig.~\ref{fig:extinction}. We observe two main tendencies: the maximal extinction (equivalent to the relative depth of the dip in the signal spectrum at $\omega_s=\omega^{(c)}=0.7$) \textit{increases} with gain, while the width of the dip \textit{decreases}.

In Fig.~\ref{fig:extinction}a, we show extinction spectra obtained at various amounts of gain, which are normalised to $1$ to better showcase the change in the spectral width as gain is increased. The lineshape of the analyte absorption spectrum (i.e. $\imag{\varepsilon_\mathrm{loss}}$) is also shown for comparison. The dependence of the extinction maximum and its full-width-at-half-maximum (FWHM) on parametric gain are shown in Fig.~\ref{fig:extinction}b. We observe that the maximum monotonically increases with gain, but with a progressively slower rate as gain rises and seems to show signs of saturation at very high gain values. On the other hand, the extinction FWHM decreases with gain (also shown in Fig.~\ref{fig:extinction}b) and shows signs of saturation at higher gain values as well. 

Both the monotonic increase of the extinction maximum and the decrease of its FWHM can be explained as a consequence of self-seeding being impeded by the analyte loss: idler photons with frequencies within the loss spectrum of the analyte are removed from the idler field before they have a chance to seed the production of further photons at those frequencies.
This results in signal photon intensity at frequencies affected by the loss to increase at a lower rate (with increasing gain) as compared to the lossless case. On the other hand, the intensity at frequencies unaffected by the loss increases at the "lossless" rate, resulting in a net increase of the maximum extinction. Analogously, the decreasing extinction FWHM with gain can also be seen as a consequence of self-seeding being impeded in the presence of loss, but here, we also need to take into account the shape of the loss spectrum. The lineshape of the analyte absorption spectrum dictates that idler photons experience less loss, the further their frequency is from the analyte resonance $\omega_\mathrm{loss}$. The idler photons further from resonance thus have a lower chance to be absorbed before seeding further pairs, leading to the signal photon intensities increasing faster (with gain) for frequencies further away from $\omega^{(c)}$, resulting in a net narrowing of the extinction spectrum.

As mentioned at the beginning of this section, sensitivity enhancement in the high-gain regime has already been observed for undetected photon measurement schemes based on induced coherence and nonlinear interferometers. The nature of the enhancement in the case of induced coherence is discussed in detail in Ref.~\cite{kolobov2017controlling} and the authors conclude that it also occurs due to seeding effects, where idler photons from the first source, after interacting with the sample, seed further pairs in the second source and thus impart the effects of the sample onto those newly generated photons.

The observed "saturation" of both the extinction maximum and FWHM at very high gain has, to our knowledge, not been reported before, but it too can be understood to be another consequence of self-seeding, more specifically, high-order self-seeding.
In such cases, idler (signal) photons of frequency $\omega$ seed the generation of further signal (idler) photons, not only at the frequency $\omega_{p0}-\omega$, but also at frequencies around it. At sufficiently high values of gain, the idler (signal) photons thus generated, can seed further signal (idler) photons and so on. This enables signal and idler photons unaffected by the loss to potentially seed the production of additional photons within the extinction spectrum. Eventually, these high-order effects compensate the gain reduction experienced by signal photons within the extinction region, resulting in the spectral shape of the extinction saturating at very high parametric gain. As these effects correspond to higher-order processes, they only appear at sufficiently high gain values ($\mathcal{G}>2$ in our case), but become more prominent as gain is further increased.

\subsection{Spatial evolution of the spectrum}
The spectra given and discussed in Sec.~\ref{sec:lossless} and \ref{sec:lossy} were obtained assuming a detector located immediately at the output of the nonlinear waveguide, however, our formalism also allows the study of the spectrum of the produced photons throughout the entire length of the nonlinear medium - all obtained by solving Eqs.~\eqref{eq:tildeABeqs} once, for a given value of parametric gain. 

\begin{figure}[htpb]
\includegraphics[width=\linewidth]{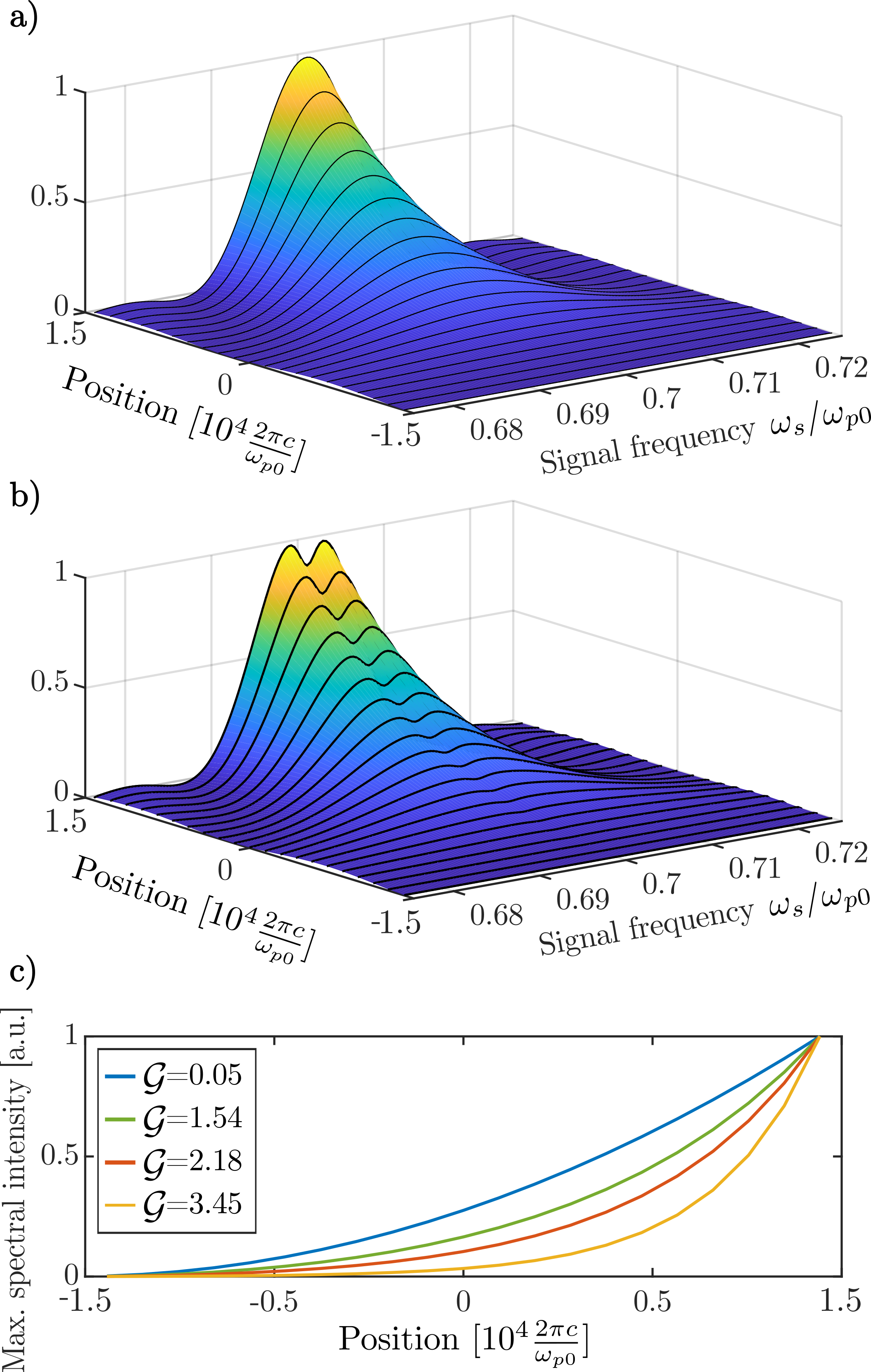}
\caption{\label{fig:3dPlot} The normalised spectral intensity of the signal photons as a function of their frequency and position within the nonlinear waveguide in the (\textbf{a}) lossless and (\textbf{b}) lossy cases. Both plots were obtained for the narrow-band pump ($\tau_p=2400$) and parametric gain of $\mathcal{G}=1.33$. (\textbf{c}) Normalised maximum spectral intensities of the signal photons as a function of position within the nonlinear waveguide, obtained for the narrow-band pump ($\tau_p=2400$) in the lossless case, for different values of gain.}
\end{figure}

Although we do not use the evolution of the spectrum within the nonlinear waveguide to obtain new insight in the context of QSUP, it can be used to showcase that our non-perturbative formalism can predict the spatial functionality of the field and field-related quantities. In Fig.~\ref{fig:3dPlot}a, we show the evolution of the signal photon spectrum as the photons propagate through the nonlinear waveguide without the analyte present. At the start of the waveguide, we see the spectrum having a very wide bandwidth and negligible amplitude; as we go forward in length, we see phase matching taking hold - reducing the bandwidth and increasing the amplitude. Additionally, in Fig.~\ref{fig:3dPlot}b we see the effect of the idler loss accumulating during propagation, causing the dip in spectral intensity, which increases with propagation, according to the discussion in Sec.~\ref{sec:lossy}. 

Finally, in Fig.~\ref{fig:3dPlot}c, we show the spatial behaviour of the spectral maximum of the signal photons for increasing values of gain, in the lossless case. We observe that, in the low-gain regime, the spectral maximum obeys the well-established quadratic law, where the intensity is $\propto l^2$, $l$ being the length of the nonlinear region. As gain is increased, the exponential nature of the length dependence of the maximum spectral intensity becomes more apparent.

\section{Conclusion}\label{sec:conclusion}
In summary, we presented a non-perturbative formalism for the description of high-gain SPDC, applicable to a wide variety of nanostructured and/or open optical systems with arbitrarily high complex spatial and spectral properties. Our formalism enables the calculation of the field operators and field-dependent quantities (e.g., correlation functions and spectra) at the output and within the nonlinear optical system.

In contrast to previous methods, our formalism is capable of describing systems with arbitrary amounts of dispersion and loss, through the use of the Green's function quantisation method, which intrinsically takes into account these effects. 

We presented both a time-domain, as well as a frequency-domain formulation, which allows the formalism to be applicable to many types of temporal and spectral analyses in the field of quantum technologies. As an example, we used this formalism to investigate quantum spectroscopy with undetected photons in a nonlinear waveguide surrounded by an analyte with a loss spectrum corresponding to the frequencies of one of the output photons. We have thus expanded upon the results of previous low-gain investigations \cite{kumar2020integrated} and discovered that the spectral sensitivity of the QSUP scheme is heavily dependent on the amount of the parametric gain and can be improved by operating the nonlinear waveguide in the high-gain regime.

Although derived here for the case of SPDC, the formalism can be generalised to treat high-gain SFWM in lossy and dispersive systems, since the SFWM interaction Hamiltonian has an identical operator structure as Eq.~\eqref{eq:hamiltonian}, when considered in the undepleted pump approximation \cite{chen2005sfwm-theory}. 

We believe the formalism presented here will advance the design and implementation of nonlinear sources of quantum light in many emerging quantum technologies, especially ones implemented on nanostructured platforms. In addition to integrated squeezed light and high-order Fock state sources in nanostructured systems with arbitrary loss and dispersion, other examples include: hybrid systems \cite{elshaari2020hybrid}, quantum sensing applications \cite{cutipa2022bsv-sensing} and quantum frequency conversion in the optical domain \cite{zaske2012visible} or microwave to optical domain \cite{han2021microwave}.

\begin{acknowledgments}
This work was supported by the Deutsche Forschungsgemeinschaft (DFG, German Research Foundation) under the project identifier 398816777-SFB 1375 (NOA), the 
German Federal Ministry of Education and Research (BMBF) under the project identifiers 13N14877 (QuantIm4Life), 13N16108 (PhoQuant), 13N15956 (LIVE2QMIC) and the 
Thuringian Ministry for Economy, Science, and Digital Society and the European Social Funds (2021 FGI 0043 -Quantum Hub Thuringia). Sina Saravi acknowledges the funding by the LIGHT profile (FSU Jena).
\end{acknowledgments}

\appendix

\section{Commutation relation for the frequency amplitude operators} \label{app:commRelation}
We begin by expanding the $\underline{\hat{E}}^{(\pm)}_{i,j}(\rr,\omega,t)$ operators in the commutator according to Eq.~\eqref{eq:gDefNormal}:
\begin{align*}
\comm{\underline{\hat{E}}_i^{(+)}(\rr,\omega,t)}{\underline{\hat{E}}_j^{(-)}(\rr',\omega',t)}&=  &\\
\mathcal{K}^2\,\frac{\omega^2}{c^2}\frac{\omega'^2}{c^2}\iint\dd{\rr_1}&\dd{\rr_2}\sqrt{\epspp(\rr_1,\omega)}\sqrt{\epspp(\rr_2,\omega')}&\\
\times& G_{ik}(\rr,\rr_1,\omega)\conj{G}_{jl}(\rr',\rr_2,\omega')&\\
\times&\comm{\opf_k(\rr_1,\omega,t)}{\opfd_l(\rr_2,\omega',t)}.&
\end{align*}
Now, we use the commutation relation Eq.~\eqref{eq:commf} and the properties of the $\delta$-functions to obtain on the right-hand side:
\begin{multline*}
\mathcal{K}^2\,\frac{\omega^4}{c^4}\delta(\omega-\omega')\\
   \times\int\dd{\rr_1} \epspp(\rr_1,\omega)G_{ik}(\rr,\rr_1,\omega)\conj{G}_{jk}(\rr',\rr_1,\omega).
\end{multline*}

The above expression can be simplified by using the following GF identity \cite{vogel2006quantum}:
\begin{eqnarray}\label{eq:gfIdentity}
    \frac{\omega^2}{c^2}\int\dd{\bar{\rr}}
    \epspp(\bar{\rr},\omega)G_{ik}(\rr,\bar{\rr},\omega)&&\conj{G}_{jk}(\rr',\bar{\rr},\omega)=\nonumber\\
    &&\imag{G_{ij}(\rr,\rr',\omega)},
\end{eqnarray}
to finally arrive at Eq.~\eqref{eq:commgNormal}.

\section{The rotating frame}\label{app:rotatingFrame}
The rotating frame operators introduced in Sec.~\ref{sec:theory1} can be more formally defined through a decomposition of the unitary evolution operator that separates the free-field evolution and the evolution induced by the nonlinear interaction. This procedure is very similar to the definition of the interaction picture of quantum mechanics.

In general, the Heisenberg equation of motion \eqref{eq:heisenbergEquation} is equivalent to the relation:
\begin{equation}\label{eq:heisPictureField}
\underline{\hat{E}}_i^{(+)}(\rr,\omega,t)=\hat{U}^\dagger(t)\underline{\hat{E}}_i^{(+)}(\rr,\omega)\hat{U}(t),
\end{equation}
where $\hat{U}(t)=\mathcal{T}\exp{\frac{1}{i\hbar}\int_{-\infty}^t\dd{t'}\left(\hat{H}_0+\hat{H}_\mathrm{SPDC}(t')\right)}$, $\hat{H}_0$ and $\hat{H}_\mathrm{SPDC}(t')$ are the free-field and nonlinear interaction Hamiltonians, respectively, both considered in the Schr\" odinger picture and $\mathcal{T}$ indicates the time-ordering superoperator. Lastly, $\underline{\hat{E}}_i^{(+)}(\rr,\omega)\equiv\underline{\hat{E}}_i^{(+)}(\rr,\omega,t\rightarrow\ -\infty)$ represents the field amplitude operators before the nonlinear interaction, where we omitted the time $t\rightarrow-\infty$ for compactness. By using Feynman's disentanglement theorem \cite{feynman1951}, we can decompose $\hat{U}(t)$ as $\hat{U}(t)=\hat{U}_0(t)\hat{U}_\mathrm{SPDC}(t)$, where:
\begin{subeqnarray}
\hat{U}_0(t)=&&\exp{\frac{1}{i\hbar}\hat{H}_0 t}\\
\hat{U}_\mathrm{SPDC}(t)=&&\mathcal{T}\exp{\frac{1}{i\hbar}\int_{-\infty}^t\dd{t'}\tilde{H}_\mathrm{SPDC}(t')}\slabel{eq:UspdcDef}\\
\tilde{H}_\mathrm{SPDC}(t)=&&\hat{U}_0^\dagger(t)\hat{H}_\mathrm{SPDC}(t)\hat{U}_0(t).\slabel{eq:HtildeDef}
\end{subeqnarray}
We can use the decomposition of $\hat{U}(t)$ to rewrite Eq.~\eqref{eq:heisPictureField} as:
\begin{equation*}
\underline{\hat{E}}_i^{(+)}(\rr,\omega,t)=\hat{U}^\dagger_\mathrm{SPDC}(t)\hat{U}_0^\dagger(t)\underline{\hat{E}}_i^{(+)}(\rr,\omega)\hat{U}_0(t)\hat{U}_\mathrm{SPDC}(t),
\end{equation*}
using $\hat{H}_0=\hbar \int\dd{\rr}\int\dd{\omega}\omega \opfd_i(\rr,\omega)\opf_i(\rr,\omega)$ and the definition \eqref{eq:gDefNormal} we then obtain:
\begin{equation*}
\underline{\hat{E}}_i^{(+)}(\rr,\omega,t)=\hat{U}^\dagger_\mathrm{SPDC}(t)\underline{\hat{E}}_i^{(+)}(\rr,\omega)
    \hat{U}_\mathrm{SPDC}(t)\exp{-i\omega t},
\end{equation*}
If we now define: 
\begin{equation}\label{eq:gDefOperator}
\opg_i^{(+)}(\rr,\omega,t)=\hat{U}^\dagger_\mathrm{SPDC}(t)\underline{\hat{E}}_i^{(+)}(\rr,\omega)
    \hat{U}_\mathrm{SPDC}(t),
\end{equation} 
we arrive at Eq.~\eqref{eq:slowlyVaryingSub}. The rotating-frame creation/annihilation operators can be defined in a similar manner to be:
\begin{equation}\label{eq:fBarDefOperator}
    \bar{f}^{(\dagger)}_i(\rr,\omega,t)=\hat{U}^\dagger_\mathrm{SPDC}(t)\hat{f}_i^{(\dagger)}(\rr,\omega)
    \hat{U}_\mathrm{SPDC}(t),
\end{equation}
where again we have $\hat{f}_i^{(\dagger)}(\rr,\omega)\equiv\hat{f}_i^{(\dagger)}(\rr,\omega,t\rightarrow-\infty)$ for compactness. As with the amplitude operators, the rotating-frame annihilation operators are related to their Heisenberg picture counterparts as:
\begin{equation}\label{eq:fBarDef}
\hat{f}_j(\rr,\omega,t)=\bar{f}_j(\rr,\omega,t)\exp{-i\omega t},
\end{equation}
with the relation for the creation operators obtained by taking the adjoint of Eq.~\eqref{eq:fBarDef}.

To find the general equation of motion for$\opg_i^{(+)}(\rr,\omega,t)$, we begin by finding the time derivative of Eq.~\eqref{eq:gDefOperator}:
\begin{eqnarray*}
\pdt \opg_i^{(+)}(\rr,\omega,t)&=&\left(\pdt \hat{U}^\dagger_\mathrm{SPDC}(t)\right)\underline{\hat{E}}_i^{(+)}(\rr,\omega)\hat{U}_\mathrm{SPDC}(t)\\
    &&+\hat{U}^\dagger_\mathrm{SPDC}(t)\underline{\hat{E}}_i^{(+)}(\rr,\omega)\left(\pdt \hat{U}_\mathrm{SPDC}(t)\right).
\end{eqnarray*}
We then recall Eq.~\eqref{eq:UspdcDef} and obtain, after some straightforward calculation:
\begin{eqnarray*}
\pdt \opg_i^{(+)}(\rr,\omega,t)=&&\\
    \frac{1}{i\hbar}\hat{U}^\dagger_\mathrm{SPDC}(t)&&\comm{\underline{\hat{E}}_i^{(+)}(\rr,\omega)}{\tilde{H}_\mathrm{SPDC}(t)}\hat{U}_\mathrm{SPDC}(t).
\end{eqnarray*}
Due to the properties of the commutator, $\hat{U}^{(\dagger)}_\mathrm{SPDC}(t)$ can act on the operators inside of it independently and give, on the right-hand side:
\begin{equation*}
    \frac{1}{i\hbar}\comm{\opg_i^{(+)}(\rr,\omega,t)}{\hat{U}^\dagger_\mathrm{SPDC}(t)\tilde{H}_\mathrm{SPDC}(t)\hat{U}_\mathrm{SPDC}(t)},
\end{equation*}\\
where we used Eq.~\eqref{eq:gDefOperator}. Also recalling Eq.~\eqref{eq:HtildeDef} and the decomposition $\hat{U}(t)=\hat{U}_0(t)\hat{U}_\mathrm{SPDC}(t)$, we can identify $\hat{U}^\dagger_\mathrm{SPDC}(t)\tilde{H}_\mathrm{SPDC}(t)\hat{U}_\mathrm{SPDC}(t)$ to be the full, Heisenberg-picture SPDC Hamiltonian given in Eq.~\eqref{eq:hamiltonian}. Thus we have proven Eq.~\eqref{eq:rotFrameEquation}.

The commutation relation for the rotating-frame creation/annihilation operators is easily found by replacing the definition Eq.~\eqref{eq:fBarDef} in the commutation relation for the Heisenberg picture operators Eq.~\eqref{eq:commf}:
\begin{equation*}
    \left[\bar{f}_i(\rr,\omega,t),\bar{f}^{\dagger}_j(\rr',\omega',t)\right]=     
    \delta_{ij}\delta(\rr-\rr')\delta(\omega-\omega')\exp{i(\omega-\omega')t}.
\end{equation*}
Since the $\delta$-function in the above expression results in $\omega\equiv\omega'$, the exponential factor can be ignored. The same is valid for the commutation relation of the rotating-frame amplitude operators, which is found by replacing Eq.~\eqref{eq:slowlyVaryingSub} in Eq.~\eqref{eq:commEbar}:
\begin{multline*}
    \comm{\opg_i^{(+)}(\rr,\omega,t)}{\opg_j^{(-)}(\rr',\omega',t)}=\\
    \mathcal{K}^2\frac{\omega^2}{c^2} \imag{G_{ij}(\rr,\rr',\omega)}\delta(\omega-\omega')\exp{i(\omega-\omega')t}.
\end{multline*}

\begin{widetext}
\section{Equations of motion for the amplitude operators}\label{app:eqsOfMotionBar}
To find the commutator $\comm{\opg_i^{(+)}(\rr,\omega,t)}{\hat{H}_\mathrm{SPDC}(t)}$, we first expand the SPDC Hamiltonian from Eq.~\eqref{eq:hamiltonian} in terms of the rotating-frame operators $\vec{\opg}^{(\pm)}(\rr,\omega,t)$:
\begin{equation*}
    \hat{H}_\mathrm{SPDC}(t)=-\varepsilon_0\int\dd{\rr'}\iint\dd{\omega'}\dd{\omega''}\chit_{ljk}(\rr') \pumpn{l}(\rr',t)
    \opg_j^{(+)}(\rr',\omega',t) \opg_k^{(+)}(\rr',\omega'',t)\exp{-i\omega't}\exp{-i\omega''t}+H.c.
\end{equation*}
Due to the linear properties of the commutator and the fact that amplitude operators with the same sign in the superscript commute, it is sufficient to calculate:
\begin{eqnarray}\label{eq:fieldsCommuting}
    \comm{\opg_{i}^{(+)}(\rr,\omega,t)}{\opg_j^{(-)}(\rr',\omega',t) \opg_k^{(-)}(\rr',\omega'',t)}= 
    \mathcal{K}^2\frac{\omega^2}{c^2}&&\Big(\delta(\omega-\omega')\imag{G_{ij}(\rr,\rr',\omega)}\opg_k^{(-)}(\rr',\omega'',t)\\
    &&+\delta(\omega-\omega'')\imag{G_{ik}(\rr,\rr',\omega)}\opg_j^{(-)}(\rr',\omega',t)\Big),
\end{eqnarray}
where we used Eq.~\eqref{eq:commEbar}. As pointed out in the main text, to simplify the relations, we assume that the Kleinman symmetry condition is valid. This is applicable when the frequency dependence of the nonlinear susceptibility is negligible, as is commonly assumed in many practical scenarios of interest \cite{boyd2020nonlinear}, as well as in the present case. In this case, we have the permutation symmetry of the nonlinear susceptibility $\chit_{ljk}=\chit_{lkj}$, hence the two terms on the right-hand-side (r.h.s.) of Eq.~\eqref{eq:fieldsCommuting} will result in identical contributions to the final equation of motion. If Kleinman symmetry is not valid, we can simply keep both terms in Eq.~\eqref{eq:fieldsCommuting}, which adds no complexity to the calculation, and only makes the expressions longer. Thus we can write:
\begin{equation*}
	\pdt \opg_{i}^{(+)}(\rr,\omega,t)=2i\frac{\varepsilon_0\mathcal{K}^2\omega^2}{\pi c^2}\int\dd{\rr'}\chit_{ljk}(\rr') \pump{l}(\rr',t)\imag{G_{ik}(\rr,\rr',\omega)}
    \iint\dd{\omega'}\dd{\omega''}\delta(\omega-\omega'')\opg_j^{(-)}(\rr',\omega',t)\exp{i\omega't}\exp{i\omega''t}.
\end{equation*}
After performing the $\omega''$-integral and replacing the value of $\mathcal{K}$, we obtain the final form of the equation of motion given in Eq.~\eqref{eq:gEqs}.

\section{Derivation of Eq.~\eqref{eq:firstOrderCorrel}}\label{app:correlationF}
To find the correlation function $g_{ij}^{(1)}(\rr,\rr';t,t')=\expval{\ne_i(\rr,t)\pe_j(\rr',t')}$ in terms of the IO-coefficients, we begin by expanding the electric field operators in the expectation value according to Eq.~\eqref{eq:EwithSlowOps}:
\begin{equation}\label{eq:productOfIOs}
	\expval{\ne_i(\rr,t)\pe_j(\rr',t')}=\iint\dd{\omega}\dd{\omega'}\expval{\opg_{i}^{(-)}(\rr,\omega,t)\opg_{j}^{(+)}(\rr',\omega',t)} \exp{i\omega t}\exp{-i\omega't'}.
\end{equation}
we then expand each of the amplitude operators using the IO-relation Eq.~\eqref{eq:gIO}. To avoid writing out all of the terms that thus appear, we will only remark that they are each proportional to one of the products: $\bar{f}_\alpha^{(\dagger)}\bar{f}_\beta^{(\dagger)}$, where $\alpha,\beta$ are arbitrary indices. The expectation value of each product is then taken in the initial state of the system (as we are working in the Heisenberg picture). In the present case, we assume the initial state to be the vacuum, which causes the expectation values to evaluate to zero for all terms except ones proportional to $\bar{f}_\alpha\bar{f}_\beta^{\dagger}$. Thus we are left with:
\begin{equation}\label{eq:differentAcoeffs}
	\expval{\ne_i(\rr,t)\pe_j(\rr',t')}=\iint\dd{\omega}\dd{\omega'}\int \dd{\XX}\dd{\XX'}\mathcal{A}_{ik}(\rr,\omega,\XX;t)\mathcal{A}^*_{jl}(\rr',\omega',\XX';t')\expval{\bar{f}_{k}(\XX)\bar{f}_l^{\dagger}(\XX')} \exp{i\omega t}\exp{-i\omega't'},
\end{equation}
where the remaining expectation value can be evaluated using the commutation relation Eq.~\eqref{eq:commFbar} to result in $\expval{\bar{f}_{k}(\XX)\bar{f}_l^{\dagger}(\XX')}=\delta_{kl}\delta(\XX-\XX')\equiv\delta_{kl}\delta(\xx-\xx')\delta(\nu-\nu')$. The $\XX'$ integral can then be evaluated to obtain Eq.~\eqref{eq:firstOrderCorrel}.

If the system was initially in a state other than vacuum, the expectation values of the various $\bar{f}_\alpha^{(\dagger)}\bar{f}_\beta^{(\dagger)}$ products may result in multiple nonzero terms, which can also be analytically evaluated using Eq.~\eqref{eq:commFbar} by first expressing the initial state using the free-field creation/annihilation operators $\bar{f}_\alpha^{(\dagger)}$.

\section{Derivation of the coupled equations for the filtered IO-coefficients} \label{app:filteredEqs}
To obtain the differential equation governing the evolution of $\tilde{A}_{ij}(\dots)$ we first take the time derivative of both sides of Eq.~\eqref{filteredABdef:1}:
\begin{equation}
    \pdt\tilde{A}_{ij}(\rr,\omega,\XX;t)=\int\dd{\omega'}\exp{-i(\omega-\omega')t}A_{ij}(\rr,\omega',\XX;t).\label{eq:AtildeDerivative}
\end{equation}
Then we write out $A_{ij}(\rr,\omega',\XX;t)$ as the formal solution of Eq.~\eqref{ABeqs:1}. This is found by integrating both sides of the equation over time:
\begin{equation*}
    A_{ij}(\rr,\omega,\XX;t)=\int_{-\infty}^t\dd{t'}S^{(0)}_{ij}(\rr,\omega,\XX;t')+\int_{-\infty}^t\dd{t'}\int \dd{\rr'}\int\dd{\omega'} \conj{F}_{ik}(\rr,\omega;\rr',\omega';t')B_{kj}(\rr',\omega',\XX;t'),
\end{equation*}
where we used the fact that $A_{ij}(\rr,\omega,\XX;t\rightarrow-\infty)=0$. When we introduce this form for $A_{ij}(\rr,\omega,\XX;t)$ into Eq.~\eqref{eq:AtildeDerivative}, we obtain two main terms which contribute to the equation of motion. The first term we label as $\tilde{S}^{(0)}_{ij}(\rr,\omega,\XX;t)$ and it has the form:
\begin{equation*}
    \tilde{S}^{(0)}_{ij}(\rr,\omega,\XX;t)=\int\dd{\omega'}\exp{-i(\omega-\omega')t}\int_{-\infty}^t\dd{t'}S^{(0)}_{ij}(\rr,\omega',\XX;t').
\end{equation*}
We can immediately rewrite it using the full expression for $S^{(0)}_{ij}(\rr,\omega',\XX;t')$, obtained from Eq.~\eqref{eq:sourceDef}:
\begin{eqnarray*}
\tilde{S}^{(0)}_{ij}(\rr,\omega,\XX;t)= \frac{2\mathcal{K}}{\pi} \frac{ \nu^2}{c^2} \sqrt{\epspp(\xx,\nu)}\int\dd{\omega'}&&\frac{\omega'^2}{c^2}\exp{-i(\omega-\omega')t}
    \int_{-\infty}^t\dd{t'}\exp{-i(\omega'+\nu)t'}\\
    &&\times\int \dd{\rr'}\chit_{klm}(\rr')\pumpn{k}(\rr',t')\imag{G_{il}(\rr,\rr',\omega')}G_{mj}(\rr',\xx,\nu).
\end{eqnarray*}

Our next aim is to be able to evaluate the time integral in the above expression. To do that, we expand the pump field into a Fourier integral:
$E_{P,k}^{(-)}(\rr,t)=\int\dd{\omega_p}\mathcal{E}_{P,k}^{(-)}(\rr,\omega_p)\exp{ i\,\omega_p t}$,
where $\mathcal{E}_P^{(-)}(\rr,\omega_p)$ is the pump amplitude in the frequency domain. The full form of $\tilde{S}^{(0)}_{ij}(\rr,\omega,\XX;t)$ is now:
\begin{eqnarray}\label{eq:tildedSourceDeriv}
    \tilde{S}^{(0)}_{ij}(\rr,\omega,\XX;t)=\frac{2\mathcal{K}}{\pi} \frac{\nu^2}{c^2} &&\sqrt{\epspp(\xx,\nu)}\int \dd{\rr'}\chit_{klm}(\rr')G_{mj}(\rr',\xx,\nu)  \nonumber \\
    &&\times
     \int\dd{\omega'}\frac{\omega'^2}{c^2}\imag{G_{il}(\rr,\rr',\omega')}\int\dd{\omega_p}\fpumpn{k}(\rr',\omega_p)\exp{-i(\omega-\omega')t}\int_{-\infty}^t\dd{t'}\exp{i(\omega_p-\omega'-\nu)t'},
\end{eqnarray}
Now we examine the time integral $\int_{-\infty}^t\dd{t'}\exp{i(\omega_p-\omega'-\nu)t'}$ and introduce a substitution $t'=t-\tau$ which allows us to rewrite it as $\exp{i(\omega_p-\omega'-\nu)t}\int_0^\infty\dd{\tau}\exp{i(\omega'+\nu-\omega_p)\tau}\equiv\exp{i(\omega_p-\omega'-\nu)t}\zeta(\omega'+\nu-\omega_p)$, where $\zeta(\omega)$ is a generalised function proportional to the Fourier transform of the Heaviside step function and is closely related to the analytical properties of the GF \cite{singh2020analytic,saravi2018photon}. One notable analytical property of the GF is \cite{saravi2018photon,andrey2016prl}:
\begin{equation}
\int\dd{\omega}\omega^2\text{Im}\left[G_{ij}(\rr,\rr',\omega)\right]\zeta(\omega-\omega_0)
=i\pi\omega_0^2\conj{G}_{ij}(\rr,\rr',\omega_0)
\end{equation}
which we can use to further simplify Eq.~\eqref{eq:tildedSourceDeriv} by performing the integration over $\omega'$. Thus we arrive at the final form for the filtered source term \eqref{eq:sourceDefTilde}.

The second term appearing on the r.h.s. of Eq.~\eqref{eq:AtildeDerivative} is:
\begin{equation*}
    \int\dd{\omega'}\exp{-i(\omega-\omega')t}\int_{-\infty}^t\dd{t'}\int \dd{\rr'}\int\dd{\omega''} \conj{F}_{ik}(\rr,\omega';\rr',\omega'';t')B_{kj}(\rr',\omega'',\XX;t').
\end{equation*}
If we write out the full form of $\conj{F}_{ik}(\rr,\omega;\rr',\omega';t')$ using Eq.~\eqref{eq:Fdefinition}, we have:
\begin{eqnarray*}
    -\frac{2i}{\pi}\int\dd{\omega'}\exp{-i(\omega-\omega')t}\int_{-\infty}^t\dd{t'}\int \dd{\rr'}\int\dd{\omega''} \frac{\omega'^2}{c^2}\chit_{lmk}(\rr')&&\imag{G_{im}(\rr,\rr',\omega')}\\
    &&\times\left(\int\dd{\omega_p}\mathcal{E}_{P,l}^{(-)}(\rr',\omega_p)\exp{ i\,\omega_p t'}\right)  
    \exp{-i(\omega'+\omega'')t'} B_{kj}(\rr',\omega'',\XX;t'),
\end{eqnarray*}
where we immediately expanded the pump field into a Fourier integral. After some reordering of the integrals in the above expression, we can identify:
\begin{equation*}
    \int_{-\infty}^t\dd{t'}\int\dd{\omega''}
\exp{i(\omega_p-\omega'-\omega'')t'}B_{kj}(\rr',\omega'',\XX;t')=\tilde{B}_{kj}(\rr',\omega_p-\omega',\XX;t),
\end{equation*}
where the r.h.s. was established according to Eq.~\eqref{filteredABdef:2}. Thus, we now write the term in question as:
\begin{equation*}
    -\frac{2i}{\pi}\int\dd{\rr'}\int\dd{\omega'}\frac{\omega'^2}{c^2}\chit_{lmk}(\rr')\imag{G_{im}(\rr,\rr',\omega')}\exp{-i(\omega-\omega')t}\int\dd{\omega_p}\fpumpn{l}(\rr',\omega_p)\tilde{B}_{kj}(\rr',\omega_p-\omega',\XX;t).
\end{equation*}
To arrive at the form of the term given by Eqs.~\eqref{tildeABeqs:1} and \eqref{eq:fDefTilde}, we first introduce a variable substitution $\omega_p=\omega'+\bar{\omega}$ into the above expression:
\begin{equation*}
    -\frac{2i}{\pi}\int\dd{\rr'}\int\dd{\omega'}\frac{\omega'^2}{c^2}\chit_{lmk}(\rr')\imag{G_{im}(\rr,\rr',\omega')}\exp{-i(\omega-\omega')t}\int_{-\omega'}^\infty\dd{\bar{\omega}}
    \fpumpn{l}(\rr',\bar{\omega}+\omega')\tilde{B}_{kj}(\rr',\bar{\omega},\XX;t).
\end{equation*}
The newly introduced $\bar{\omega}$ has the limits $[-\omega',\infty)$, as shown above, which indicates that the filtered coefficients have to be evaluated at certain negative frequency values during calculation. Although this is also formally allowed by the definitions in Eqs.~\eqref{eq:filteredABdef}, these negative-frequency contributions can be safely neglected and the lower integration limit set to $0$ without any loss of accuracy. The reason for this can be seen by examining the source term in \eqref{eq:sourceDefTilde}, where negative values of the variable $\omega$ would cause the term to oscillate at optical frequencies, thus averaging to $0$ on the time scales of the SPDC process and resulting in these negative-frequency values of the IO-coefficients not contributing in the final equations of motion. Finally, to make the resulting expression consistent with the notation used in the main text, we will exchange the variables $\omega'$ and $\bar{\omega}$ and write the final form of the second term of Eq.~\eqref{eq:AtildeDerivative}:
\begin{equation*}
    -\frac{2i}{\pi}\int\dd{\rr'}\int\dd{\omega'}\left(\int\dd{\bar{\omega}}\frac{\bar{\omega}^2}{c^2}\chit_{lmk}(\rr')\imag{G_{im}(\rr,\rr',\bar{\omega})}\exp{-i(\omega-\bar{\omega})t}
    \fpumpn{l}(\rr',\omega'+\bar{\omega})\right)\tilde{B}_{kj}(\rr',\omega',\XX;t),
\end{equation*}
where the expression in the parentheses exactly corresponds to the definition Eq.~\eqref{eq:fDefTilde} and, when the above expression is combined with Eq.~\eqref{eq:sourceDefTilde}, we get exactly the equation of motion Eq.~\eqref{tildeABeqs:1}.
An analogous procedure can be performed to find the equation of motion for the coefficient $\tilde{B}_{kj}(\rr,\omega,\XX;t)$, after which we obtain the coupled equations \eqref{eq:tildeABeqs}.

\section{Low-gain solution}\label{app:perturbative}

To test the validity of the coupled equations Eqs.~\eqref{eq:tildeABeqs}, we find the perturbative solutions for the IO-coefficients in the the low-gain regime, when the pump amplitude is sufficiently weak, and use them to calculate the low-gain single-photon spectrum.

The single-photon output spectrum in terms of the IO-coefficients is given in Eq.~\eqref{eq:probWithAtilde}, where we see that it is entirely determined by $\tilde{A}_{kj}(\rr,\omega,\XX;t)$. Thus, an analytical expression for the spectrum in the low-gain regime can be obtained by finding a first-order perturbative solution for the IO-coefficient $\tilde{A}_{kj}(\rr,\omega,\XX;t)$ and replacing it in Eq.~\eqref{eq:probWithAtilde}.

To obtain a perturbative expansion of $\tilde{A}_{kj}(\rr,\omega,\XX;t)$, we begin by integrating both sides of  Eq.~\eqref{tildeABeqs:1} over time, to find the formal form of the solution. This yields:
\begin{equation}\label{eq:formalAtilde}
    \tilde{A}_{ij}(\rr,\omega,\XX;t)=\int_{-\infty}^t\dd{t'}\tilde{S}^{(0)}_{ij}(\rr,\omega,\XX;t')+\int_{-\infty}^t\dd{t'}\int \dd{\rr'}\int\dd{\omega'} \conj{\tilde{F}}_{ik}(\rr,\omega;\rr',\omega';t')\tilde{B}_{kj}(\rr',\omega',\XX;t').
\end{equation}
Next, we find the formal solution of Eq.~\eqref{tildeABeqs:2} in the same manner:
\begin{equation*}
    \tilde{B}_{ij}(\rr,\omega,\XX;t)=\int_{-\infty}^t\dd{t'}\int \dd{\rr'}\int\dd{\omega'} \tilde{F}_{ik}(\rr,\omega;\rr',\omega';t')\tilde{A}_{kj}(\rr',\omega',\XX;t'),
\end{equation*}
and insert it into \eqref{eq:formalAtilde}:
\begin{eqnarray}\label{eq:formalAtildeExpansion}
    &&\tilde{A}_{ij}(\rr,\omega,\XX;t)=\int_{-\infty}^t\dd{t'}\tilde{S}^{(0)}_{ij}(\rr,\omega,\XX;t')+\\
    &&\qquad\int_{-\infty}^t\dd{t'}\int \dd{\rr'}\int\dd{\omega'} 
    \conj{\tilde{F}}_{ik}(\rr,\omega;\rr',\omega';t')
    \left(\int_{-\infty}^{t'}\dd{t''}\int \dd{\rr''}\int\dd{\omega''} \tilde{F}_{kl}(\rr',\omega';\rr'',\omega'';t'')\tilde{A}_{lj}(\rr'',\omega'',\XX;t'')\right).
\end{eqnarray}
The full perturbative expansion is obtained by iteratively expanding the above expression using the formal solutions of Eqs.~\eqref{eq:tildeABeqs}. However, as we are interested in the \textit{first-order} solution for $\tilde{A}_{ij}(\rr,\omega,\XX;t)$, Eq.~\eqref{eq:formalAtildeExpansion} is already sufficient. This can be confirmed by noting that both $\tilde{S}^{(0)}_{ij}(\rr,\omega,\XX;t)$ and $\tilde{F}_{ik}(\rr,\omega;\rr',\omega';t)$ are proportional to the pump amplitude, according to Eqs.~\eqref{eq:sourceDefTilde} and \eqref{eq:fDefTilde}, respectively, and any further expansion will result in terms of third- and higher-order in the pump amplitude.

Thus we conclude that the first-order analytical expression for $\tilde{A}_{ij}(\rr,\omega,\XX;t)$, is just the source term $\tilde{S}^{(0)}_{ij}(\rr,\omega,\XX;t)$ integrated over time. Using Eq.~\eqref{eq:sourceDefTilde} we find it has the form:
\begin{equation}\label{eq:Apert}
    \tilde{A}_{ij}(\rr,\omega,\XX;t\rightarrow\infty)=4i\pi\mathcal{K}\, \frac{\nu^2\omega^2}{c^2} \sqrt{\epspp(\xx,\nu)}
    \int \dd{\rr'}\chit_{klm}(\rr')\fpumpn{k}(\rr',\omega+\nu)
    \conj{G}_{il}(\rr,\rr',\omega)G_{mj}(\rr',\xx,\nu).
\end{equation}
Now we insert the above expression into \eqref{eq:probWithAtilde} to obtain the spectrum in the low-gain regime. The resulting expression is rather long but, through some straightforward algebra and an application of Eq.~\eqref{eq:gfIdentity}, it can be reduced to obtain:
\begin{eqnarray}\label{eq:probPert}
\sigma(\omega_0,t\rightarrow\infty)\propto \frac{\omega_0^4}{c^4} \iint\dd{\rr'}\dd{\rr''}&&\chit_{klm}(\rr')\chit_{qrs}(\rr'')G_{il}^*(\rr_0,\rr',\omega_0)G_{ir}(\rr_0,\rr'',\omega_0) \nonumber \\
\times &&\int\dd{\nu}\frac{\nu^2}{c^2}\fpumpn{k}(\rr',\omega_0+\nu)\fpump{q}(\rr'',\omega_0+\nu)\imag{G_{ms}(\rr',\rr'',\nu)},
\end{eqnarray}
 \end{widetext}

\section{Notes on numerical implementation}\label{app:aux}
In both the time- and frequency-domain formulations, the relatively high dimensionality of the IO-coefficients $\tilde{A}_{ij}(\rr,\omega,\XX;t)$ and $\tilde{B}_{ij}(\rr,\omega,\XX;t)$ requires a careful approach to discretising the relevant ranges for each of the variables present in order to minimise the computational resources required for solving the coupled equations Eq.~\eqref{eq:tildeABeqs}.
In the case of the QSUP scheme investigated in this work, the one-dimensional nature of the waveguide and the homogeneity of the waveguide dielectric result in a phase matching function that is well-approximated by $\text{sinc}^2\left(\Delta k(\omega_p,\omega_s,\omega_i)\frac{L}{2}\right)$, where $\omega_p$, $\omega_s$ and $\omega_i$ are the pump, signal and idler frequencies, respectively and $\Delta k(\omega_p,\omega_s,\omega_i)$ is the phase mismatch. With this in mind, the width of the frequency range of interest $\Delta \omega$ was chosen to include frequencies around the central and first-order peaks of the $\text{sinc}$-function for both the signal and idler photons. The spatial discretisation step $\delta z$ was then chosen to satisfy $\delta z\ll \frac{L}{2 \pi }$ to ensure that phase-matching is correctly reproduced.

The ranges and discretisation steps for the variables $\xx$ and $\nu$ are determined in a more involved manner, but one that is entirely defined by the properties of the nonlinear waveguide and the behaviour of its GF. As noted in Sec.~\ref{sec:IOrelations}, $\xx$ and $\nu$ serve to index the initial-time field-matter excitations taking part in the IO-relations \eqref{eq:IOsourced} and \eqref{eq:filteredIO} with the contributions from each excitation being weighted by the IO-coefficients. In theory, all possible values for these variables should be taken into account in order to obtain the exact output fields, however, as will now be discussed, the relevant contributions come from specific limited ranges determined by the spatial and spectral properties of the structure in question, as well as the nonlinear interaction.

To determine which range of the spatial variable $\xx$ gives a dominant contribution to the output we examine the low-gain analytical expression for the coefficient $\tilde{A}(z,\omega,\XX;t\rightarrow\infty)$ (since it is the only one contributing to the spectrum) given by simplifying Eq.~\eqref{eq:Apert} according to the assumptions made about the waveguide under consideration in Section~\ref{sec:qsup}:
\begin{eqnarray}\label{eq:1dApert}
    \tilde{A}(z,\omega,\XX;t\rightarrow\infty)&=&2i\mathcal{K}\, \frac{\nu^2\omega^2}{c^2} \sqrt{\epspp(\nu)}\nonumber \\
    &&\times\int \dd{z'}\chit(z')\fpumpn{}(z',\omega+\nu)\qquad\nonumber \\
    &&\times\conj{G}(z,z',\omega)G(z',\xi,\nu).\qquad
\end{eqnarray}
The only $\xi$-dependent factor present - $G(z',\xi,\nu)$, in the case of the GF given by \eqref{eq:1dGF}, is proportional to $\exp{i\frac{\nu}{c}\sqrt{\varepsilon(\nu)}\abs{z'-\xi}}$. Due to $\varepsilon(\nu)$ being complex, this factor represents an oscillating function with an exponentially decreasing amplitude in $\abs{z'-\xi}$ with the decay length $l_{d}(\nu)=\frac{1}{\frac{\nu}{c}\imag{\sqrt{\varepsilon(\nu)}}}$. Since the variable $z'$ is always confined to the nonlinear region of the waveguide, the effective range of the variable $\xi$ consists of the nonlinear region itself and a multiple of the length $l_d$ of the linear regions surrounding the nonlinear waveguide. Physically this means that field-matter excitations sufficiently far away from the nonlinear region are unable to influence the nonlinear interaction due to their effects being damped away by the absorption of the dielectric. 

In practice, it is sufficient to find the largest possible decay length for the frequency range of interest $l_{d,max}$ and fix that as the boundary value for $\xi$. For materials with small amounts of loss or where the loss is very narrow-band, these boundary values can be extremely large and could present a problem in a numerical implementation, where memory limitations are a factor.

The potentially large range of $\xi$ outside of the nonlinear region can be made more manageable via a variable transformation. We used the double-exponential transformation of the form \cite{mori2001doubleExp}: $\xi=\sinh(\frac{\pi}{2}\sinh{\theta})$ where $\theta$ is the substitution variable whose range is chosen to correspond to values of $\xi$ outside the nonlinear region. With such a transformation, 
we compress the potentially long range of integration for the variable $\xi$ in the surrounding linear regions into a much shorter range for the variable $\theta$.
This transformation resembles the coordinate transformation used in implementing perfectly matched layers (PMLs) in computational photonics \cite{sauvan2022normalization}.

Even though the above conclusions were made using the low-gain result, they also hold in the high-gain regime as well, as was confirmed by our simulations. This is due to the spatial/absorbing properties of the structure being independent of the pump intensity and thus, the spatial distribution of the initial-time field-matter excitations contributing to the output stays the same, regardless of gain.

On the other hand, the relevant frequency range for the contributing initial-time field-matter excitations, i.e. the relevant range of $\nu$, cannot simply be inferred from the low-gain, due to the seeding effect that couples neighbouring frequencies in the high-gain regime. Looking again at Eq.~\eqref{eq:probPert}, we observe the following in the low-gain regime: the range of frequencies $\nu$ contributing to the output at frequency $\omega_0$ is determined by the spectrum of the pump, i.e. output photons of frequency $\omega_0$ are influenced by field-matter excitations of frequencies within a pump bandwidth of $\nu_0=\omega_{p0}-\omega_0$. 
However, as gain is increasing, the generated field at frequencies neighbouring $\omega_0$ will start contributing to the intensity at $\omega_0$ through higher order seeding effects, where those neighbouring frequencies themselves are affected by their own respective neighbouring frequencies further from $\omega_0$. Hence, to have an accurate representation of the intensity at $\omega_0$, one has to include the whole range of frequencies with substantial intensities around $\omega_0$, which are mainly determined by the phase-matching condition. Hence, in our calculation, for both $\omega$ and $\nu$ that appear in Eqs.~\eqref{eq:tildeABeqs}, we consider the whole range of frequencies around the main and the first surrounding peaks of the phase-matching function.

\bibliography{bibliography}

\end{document}